\documentclass[12pt]{iopart}
\usepackage{graphicx}

\usepackage{bm}

\usepackage{epsfig}

\usepackage{makeidx} 
\usepackage{wrapfig}
\usepackage{float}
\usepackage{pstricks,pst-grad,pst-plot}
\usepackage{graphicx}
\usepackage{dcolumn}
\usepackage{epsfig}
\usepackage{verbatim}
\usepackage{pst-grad,pst-node,pst-tree,pst-plot}
\usepackage{pstricks}

\begin{document}

\title{
Exact spin-spin correlation function for 
the zero-temperature 
random-field Ising model}

\author{T P Handford$^1$,
        F-J Perez-Reche$^2$ and
        S N Taraskin$^1$}

\address{$^1$ Department of Chemistry, 
              University of Cambridge, Cambridge, UK}
\ead{\mailto{tph32@cam.ac.uk}, \mailto{snt1000@cam.ac.uk}}          

\address{$^2$ SIMBIOS Centre, University of Abertay Dundee,  Dundee,  UK}
\ead{p.perezReche@abertay.ac.uk}

\date{Received: date / Accepted: date}

\begin{abstract}
An exact expression for the spin-spin correlation function is derived for the zero-temperature random-field Ising model defined on a Bethe lattice of arbitrary coordination number.
The correlation length describing dynamic spin-spin correlations and separated from the intrinsic topological length scale of the Bethe lattice is shown to diverge as a power law at the critical point. 
The critical exponents governing the behaviour of the correlation length are consistent with the mean-field values found for a hypercubic lattice with dimension greater than the upper critical dimension.
\end{abstract}

\maketitle 

\section{Introduction}
\label{sec:Intro}

The zero-temperature random-field Ising model (zt-RFIM) is a prototype model for systems that exhibit avalanche dynamics when slowly driven accross athermal first-order phase transitions~\cite{Sethna1993,Sethna2001}. 
Examples of such transitions include the condensation of fluids in porous media~\cite{Lilly1993}, the martensitic transformation~\cite{Rosinberg_Vives_Review2010}, or magnetisation reversal of ferromagnets (Barkhausen effect)~\cite{Durin_review2004}. 
For experimentally reasonable time scales, thermal fluctuations do not play an important role and the dynamics in this kind of system proceed along a non-equilibrium path consisting of metastable states~\cite{PerezReche2001}. Avalanches are a manifestation of such behaviour and correspond to the driven-induced passage between two metastable states. The zt-RFIM predicts that the properties of avalanches (e.g. their size or duration) are drastically affected by the degree of quenched disorder in the system. More explicitly, the model predicts three different regimes for avalanche behaviour depending on the degree of disorder. 
The main feature of the small-disorder regime is the existence of an avalanche that is infinite in extent. In contrast, high degrees of disorder lead to a regime where all avalanches are small. These two regimes are separated by an intermediate situation where the model exhibits critical behaviour (i.e. a continuous phase transition)~\cite{Sethna1993,Dahmen1993,Dhar1997,Perkovic1999}.
For any degree of disorder, avalanche-like dynamics lead to spacial correlations in observable properties averaged over quenched disorder (e.g. magnetisation, stress or fluid density), which extend over a typical length scale, called the correlation length.
This quantity is expected to diverge in the critical regime, and this has been shown to be the case in the systems described by mean field~\cite{Dahmen1993} and hypercubic lattices~\cite{Perkovic1999}.
Concerning other topologies, including a Bethe lattice, the situation is less clear.

Previous studies of the zt-RFIM on the Bethe lattice have obtained exact results for the magnetisation~\cite{Dhar1997}, the avalanche size distribution~\cite{Sabhapandit2000}, the different contributions to the energy~\cite{Illa_PRB2005_EnergyRFIM}, and the number of metastable states~\cite{Detcheverry2005,PerezReche_PRB2008,Rosinberg_JSTAT2009}.
In particular, it has been demonstrated that for coordination numbers $q>3$ the system exhibits a discontinuity in its magnetisation hysteresis loop for small amounts of disorder associated with the infinite avalanche.
The universality class of the critical point on such a lattice has been suggested to be the same as that of the mean-field system (complete graph)~\cite{Sethna1993,Dahmen1993}, with the critical exponents for the 
order parameter, $\beta=1/2$ and $\delta=3$, being identical to the mean-field values~\cite{Dhar1997,Illa_PRB2006_BetheMF}. 
The spatial correlations of the model have also been investigated in the past but mostly using numerical simulations~\cite{Perkovic1999} or approximate analytical methods based on mean-field descriptions and/or renormalization group (RG) techniques~\cite{Dahmen1996}.
In this paper, we derive an exact analytical expression for the spin-spin correlation function corresponding to the zt-RFIM with spins placed on a Bethe lattice. Our results are exact for any coordination number and confirm the validity of the functional form for the correlation function derived independently in~\cite{Rosinberg2011}, where it is assumed that the extension from the one-dimensional case (i.e. with $q=2$) is valid.

It is well established that in a hypercubic lattice of dimensionality $d$ around criticality, the correlation function, $C(r)$, decays exponentially with distance $r$ for the zt-RFIM, 
\begin{equation}
C(r)=A(r)\exp(-r/\xi)~,
\label{eq:CorrFunc1}
\end{equation}
where $A(r)$ obeys a power law for large $r$, and $\xi$ is the correlation length.
At the critical point, $\xi \to \infty$, and so $A(r)$ represents the critical behaviour of the correlation function, found to be 
\[
A(r)\propto 1/r^{d-4+\bar{\eta}}~, 
\]
with anomalous dimension $\bar{\eta}$~\cite{Dahmen1993,Dahmen1996}.
In a prototype loopless topology, the Cayley tree or Bethe lattice, the correlation function exhibits additional exponential behaviour due to the associated hyperbolic topology, 
\begin{equation}
C(r)=B(r)(q-1)^{-r}\exp(-r/\xi)~, 
\label{eq:CorrFunc2}
\end{equation}
where $q$ is the coordination number and $B(r)$ is a power-law function for large values of chemical distance $r$.
The exponential prefactor $(q-1)^{r}$ in~\eref{eq:CorrFunc2} gives the number of lattice sites within $r$ shells of the Bethe lattice, and plays the role of the factor $r^{d}$ in the equivalent expression for hypercubic lattices.
The function 
\[
B(r)\propto 1/r^{\tilde{\eta}}~,
\]
with some exponent $\tilde{\eta}$ (found below to be $\tilde{\eta}=-1$), accounts for all power-law behaviour of the correlation function.
There is some inconsistency in the definition of the correlation length for two-state models (the equilibrium Ising model, the zt-RFIM and percolation) on the Bethe lattice.
In the equilibrium Ising model~\cite{Mukamel1974} and the zt-RFIM~\cite{Rosinberg2011} on the Bethe lattice, the prefactor $(q-1)^{-r}$ in~\eref{eq:CorrFunc2} has been absorbed into the definition of the correlation length 
\begin{equation}
C(r)\propto(q-1)^{-r}\exp(-r/\xi)=\exp(-r/r_0)~,
\label{eq:CorrLen1}
\end{equation}
with 
\begin{equation}
r_0=(\ln(q-1)+\xi^{-1})^{-1}~,
\label{eq:r0}
\end{equation}
chosen to be the correlation length.
In an alternative definition, for both percolation~\cite{Straley1982} and the Ising model~\cite{Hu1997}, the correlation length has instead been identified with $\xi$ in~\eref{eq:CorrFunc2}, thus separating the length scale of dynamic correlations from the intrinsic topological length scale of the Bethe lattice, $1/\ln(q-1)$.

Below, we demonstrate that (i) the value of $\xi$ diverges at the critical point according to a power law, in contrast to $r_0$ which remains finite at criticality and (ii) the critical exponents governing the divergence of $\xi$ around the critical point are consistent with the mean-field values for the divergence of correlation length in hypercubic lattices above the upper critical dimension, $d\geq d_c=6$.
The exponents, describing the divergence of $\xi$ in terms of the external field $H$ and degree of disorder $\Delta$, $\xi\propto (H-H_c)^{-\mu}$ and $\xi\propto (\Delta-\Delta_c)^{-\nu}$, are found to be $\mu=2/3$ and $\nu=1$.
As argued below, these values can be related to those in the mean-field model, $\mu/\bar{d}=\mu_{\rm{MF}}=1/3$ and $\nu/\bar{d}=\nu_{\rm{MF}}=1/2$, where $\bar{d}=2$ originates from fractal dimension of a random walk~\cite{Straley1982,deGennes1976}.
Therefore these findings motivate us to choose $\xi$ as the correlation length describing the critical behaviour of the zt-RFIM on a Bethe lattice.
Such a definition, while in contrast with that used in~\cite{Rosinberg2011}, separates the topological contribution of the Bethe lattice to the correlation length, and allows its comparison with the correlation length for hypercubic lattices.

\section{Model}
\label{sec:Model}

The zt-RFIM involves a set of $N$ Ising spins, $\left\{ s_i=\pm 1 | i=1,2,\ldots, N \right\}$, interacting ferromagnetically 
with strength $J>0$, affected by an external field $H$ and quenched local disorder, manifested in independent random local fields, $h_i$, at each site $i$, according to the following Hamiltonian,
\begin{equation}
{\cal H}=-J\sum_{\langle ij\rangle}s_is_j-H \sum_is_i-\sum_ih_is_i~,\label{eq:Hamiltonian}
\end{equation}
where $\langle ij\rangle$ denotes a sum over all pairs of nearest neighbours.
The random fields $h_i$ are assumed to be identically distributed according to the probability density function $\rho(h_i)$ with, for convenience, zero mean and standard deviation $\Delta$.
It has been found that in certain topologies of the network of spins, and for disorder $\Delta<\Delta_c$, the zt-RFIM exhibits a discontinuity in the hysteresis loop for magnetisation, $m=N^{-1}\sum_is_i$, at some coersive field $H(\Delta)$, where the magnetisation in a large fraction of the system reverses in a single avalanche.
Increasing disorder is found to reduce the size of this discontinuity, and to remove it continuously at the critical value of disorder $\Delta_c$, around which spin-spin correlations become infinite-ranged and the system exhibits scale-free universal properties~\cite{Sethna1993,Dhar1997,Perkovic1999,PerezReche2004RFIMField}. 

We start with a description of the dynamical rules governing the relaxation of spins within the zt-RFIM. 
The external field is initially set to $H=-\infty$, forcing all spins to be $s_i=-1$, and then allowed to increase adiabatically to $H=\infty$.
When the external field is varied, the system becomes unstable and relaxes into a new metastable state 
through a series of spin flips.
We assume that relaxation takes place according to Glauber single-spin flip dynamics for the zero temperature case~\cite{Dhar1997}, such that a spin flips only if such a flip reduces the 
overall energy. 
If a spin at node $i$ changes state (i.e. $s_i \rightarrow -s_i$), it induces a change in energy $2f_i
  s_i$, where $f_i= J\sum_{\langle i|j\rangle}s_j+H+h_i$ is a local
  field. 
If $f_is_i<0$ then the spin flip will occur, otherwise it remains aligned with the local field.
As spins flip one at a time~\cite{Sethna1993,Dhar1997}, 
the local field at surrounding sites changes, and other spins may become unstable and also flip, therefore relaxation occurs in an avalanche like manner.
At a given external field, this process continues until all spins are aligned with their
respective local fields at which point the system becomes stable, thus creating clusters of flipped spins around each nucleation site. 
As the external field increases avalanches continue to progress, 
and it is known that 
the set of spins that flip in multiple avalanches when the external field is slowly increased from $-\infty$ to $H$ coincides with the set of spins that would flip if the field had stepped from $-\infty$ to $H$ (see~\cite{Dhar1997}). 
Accordingly, it can be assumed that the correlation function calculated at any external field depends only on the properties of avalanches occurring at that field.

\begin{figure}
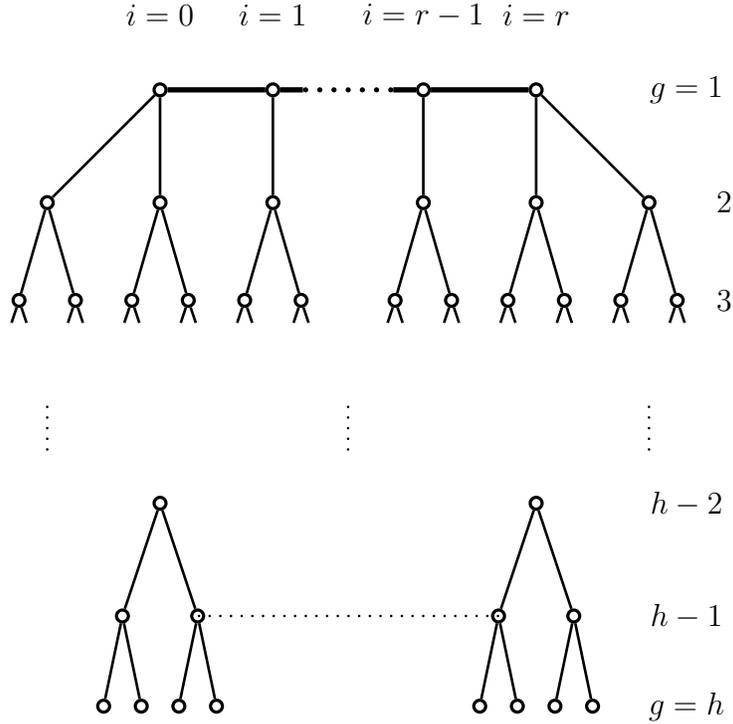

\vskip1.5truecm
\begin{center}
\psset{unit=1.0cm}
\rput(4.5,0.0){ $g=1$ }
\rput(-2.5,1.0){ $i=0$ }
\cnode[linewidth=1.2pt](-2.5,0.0){0.1}{a1}
\rput(-1.0,1.0){ $i=1$ }
\cnode[linewidth=1.2pt](-1.0,0.0){0.1}{a2}
\cnode[linewidth=0pt](-0.6,0.0){0}{a2p1}
\cnode[linewidth=0pt](0.6,0.0){0}{a2p2}
\rput(1.0,1.0){ $i=r-1$ }
\cnode[linewidth=1.2pt](1.0,0.0){0.1}{a3}
\rput(2.5,1.0){ $i=r$ }
\cnode[linewidth=1.2pt](2.5,0.0){0.1}{a4}
\ncline[linewidth=2.0pt]{a1}{a2}
\ncline[linewidth=2.0pt]{a2}{a2p1}
\ncline[linewidth=2.0pt,linestyle=dotted]{a2p1}{a2p2}
\ncline[linewidth=2.0pt]{a2p2}{a3}
\ncline[linewidth=2.0pt]{a3}{a4}
\rput(5,-1.5){ $2$ }
\cnode[linewidth=1.2pt](-4.0,-1.5){0.1}{b1}
\cnode[linewidth=1.2pt](-2.5,-1.5){0.1}{b2}
\cnode[linewidth=1.2pt](-1.0,-1.5){0.1}{b3}
\cnode[linewidth=1.2pt](1.0,-1.5){0.1}{b4}
\cnode[linewidth=1.2pt](2.5,-1.5){0.1}{b5}
\cnode[linewidth=1.2pt](4.0,-1.5){0.1}{b6}
\ncline[linewidth=1.0pt]{a1}{b1}
\ncline[linewidth=1.0pt]{a1}{b2}
\ncline[linewidth=1.0pt]{a2}{b3}
\ncline[linewidth=1.0pt]{a3}{b4}
\ncline[linewidth=1.0pt]{a4}{b5}
\ncline[linewidth=1.0pt]{a4}{b6}
\rput(5,-2.8){ $3$ }
\cnode[linewidth=1.2pt](-4.375,-2.8){0.1}{c1}
\cnode[linewidth=1.2pt](-3.625,-2.8){0.1}{c2}
\cnode[linewidth=1.2pt](-2.875,-2.8){0.1}{c3}
\cnode[linewidth=1.2pt](-2.125,-2.8){0.1}{c4}
\cnode[linewidth=1.2pt](-1.375,-2.8){0.1}{c5}
\cnode[linewidth=1.2pt](-0.625,-2.8){0.1}{c6}
\cnode[linewidth=1.2pt](0.625,-2.8){0.1}{c7}
\cnode[linewidth=1.2pt](1.375,-2.8){0.1}{c8}
\cnode[linewidth=1.2pt](2.125,-2.8){0.1}{c9}
\cnode[linewidth=1.2pt](2.875,-2.8){0.1}{c10}
\cnode[linewidth=1.2pt](3.625,-2.8){0.1}{c11}
\cnode[linewidth=1.2pt](4.375,-2.8){0.1}{c12}
\ncline[linewidth=1.0pt]{b1}{c1}
\ncline[linewidth=1.0pt]{b1}{c2}
\ncline[linewidth=1.0pt]{b2}{c3}
\ncline[linewidth=1.0pt]{b2}{c4}
\ncline[linewidth=1.0pt]{b3}{c5}
\ncline[linewidth=1.0pt]{b3}{c6}
\ncline[linewidth=1.0pt]{b4}{c7}
\ncline[linewidth=1.0pt]{b4}{c8}
\ncline[linewidth=1.0pt]{b5}{c9}
\ncline[linewidth=1.0pt]{b5}{c10}
\ncline[linewidth=1.0pt]{b6}{c11}
\ncline[linewidth=1.0pt]{b6}{c12}
\cnode[linewidth=0pt](-4.275,-3.1){0.0}{c1p1}
\cnode[linewidth=0pt](-3.525,-3.1){0.0}{c2p1}
\cnode[linewidth=0pt](-2.775,-3.1){0.0}{c3p1}
\cnode[linewidth=0pt](-2.025,-3.1){0.0}{c4p1}
\cnode[linewidth=0pt](-1.275,-3.1){0.0}{c5p1}
\cnode[linewidth=0pt](-0.525,-3.1){0.0}{c6p1}
\cnode[linewidth=0pt](0.525,-3.1){0.0}{c7p1}
\cnode[linewidth=0pt](1.275,-3.1){0.0}{c8p1}
\cnode[linewidth=0pt](2.025,-3.1){0.0}{c9p1}
\cnode[linewidth=0pt](2.775,-3.1){0.0}{c10p1}
\cnode[linewidth=0pt](3.525,-3.1){0.0}{c11p1}
\cnode[linewidth=0pt](4.275,-3.1){0.0}{c12p1}
\cnode[linewidth=0pt](-4.475,-3.1){0.0}{c1p2}
\cnode[linewidth=0pt](-3.725,-3.1){0.0}{c2p2}
\cnode[linewidth=0pt](-2.975,-3.1){0.0}{c3p2}
\cnode[linewidth=0pt](-2.225,-3.1){0.0}{c4p2}
\cnode[linewidth=0pt](-1.475,-3.1){0.0}{c5p2}
\cnode[linewidth=0pt](-0.725,-3.1){0.0}{c6p2}
\cnode[linewidth=0pt](0.725,-3.1){0.0}{c7p2}
\cnode[linewidth=0pt](1.475,-3.1){0.0}{c8p2}
\cnode[linewidth=0pt](2.225,-3.1){0.0}{c9p2}
\cnode[linewidth=0pt](2.975,-3.1){0.0}{c10p2}
\cnode[linewidth=0pt](3.725,-3.1){0.0}{c11p2}
\cnode[linewidth=0pt](4.475,-3.1){0.0}{c12p2}
\ncline[linewidth=1.0pt]{c1}{c1p1}
\ncline[linewidth=1.0pt]{c2}{c2p1}
\ncline[linewidth=1.0pt]{c3}{c3p1}
\ncline[linewidth=1.0pt]{c4}{c4p1}
\ncline[linewidth=1.0pt]{c5}{c5p1}
\ncline[linewidth=1.0pt]{c6}{c6p1}
\ncline[linewidth=1.0pt]{c7}{c7p1}
\ncline[linewidth=1.0pt]{c8}{c8p1}
\ncline[linewidth=1.0pt]{c9}{c9p1}
\ncline[linewidth=1.0pt]{c10}{c10p1}
\ncline[linewidth=1.0pt]{c11}{c11p1}
\ncline[linewidth=1.0pt]{c12}{c12p1}
\ncline[linewidth=1.0pt]{c1}{c1p2}
\ncline[linewidth=1.0pt]{c2}{c2p2}
\ncline[linewidth=1.0pt]{c3}{c3p2}
\ncline[linewidth=1.0pt]{c4}{c4p2}
\ncline[linewidth=1.0pt]{c5}{c5p2}
\ncline[linewidth=1.0pt]{c6}{c6p2}
\ncline[linewidth=1.0pt]{c7}{c7p2}
\ncline[linewidth=1.0pt]{c8}{c8p2}
\ncline[linewidth=1.0pt]{c9}{c9p2}
\ncline[linewidth=1.0pt]{c10}{c10p2}
\ncline[linewidth=1.0pt]{c11}{c11p2}
\ncline[linewidth=1.0pt]{c12}{c12p2}
\psline[linewidth=1.0pt,linestyle=dotted](-4.0,-4.2)(-4.0,-4.8)
\psline[linewidth=1.0pt,linestyle=dotted](0.0,-4.2)(0.0,-4.8)
\psline[linewidth=1.0pt,linestyle=dotted](4.0,-4.2)(4.0,-4.8)
\rput(4.5,-5.5){ $h-2$ }
\cnode[linewidth=1.2pt](-2.5,-5.5){0.1}{d1}
\cnode[linewidth=1.2pt](2.5,-5.5){0.1}{d2}
\rput(4.5,-7.0){ $h-1$ }
\cnode[linewidth=1.2pt](-3.0,-7.0){0.1}{e1}
\cnode[linewidth=1.2pt](-2.0,-7.0){0.1}{e2}
\cnode[linewidth=1.2pt](2.0,-7.0){0.1}{e3}
\cnode[linewidth=1.2pt](3.0,-7.0){0.1}{e4}
\ncline[linewidth=1.0pt]{d1}{e1}
\ncline[linewidth=1.0pt]{d1}{e2}
\ncline[linewidth=1.0pt]{d2}{e3}
\ncline[linewidth=1.0pt]{d2}{e4}
\rput(4.5,-8.2){ $g=h$ }
\cnode[linewidth=1.2pt](-3.25,-8.2){0.1}{f1}
\cnode[linewidth=1.2pt](-2.75,-8.2){0.1}{f2}
\cnode[linewidth=1.2pt](-2.25,-8.2){0.1}{f3}
\cnode[linewidth=1.2pt](-1.75,-8.2){0.1}{f4}
\cnode[linewidth=1.2pt](1.75,-8.2){0.1}{f5}
\cnode[linewidth=1.2pt](2.25,-8.2){0.1}{f6}
\cnode[linewidth=1.2pt](2.75,-8.2){0.1}{f7}
\cnode[linewidth=1.2pt](3.25,-8.2){0.1}{f8}
\ncline[linewidth=1.0pt]{e1}{f1}
\ncline[linewidth=1.0pt]{e1}{f2}
\ncline[linewidth=1.0pt]{e2}{f3}
\ncline[linewidth=1.0pt]{e2}{f4}
\ncline[linewidth=1.0pt]{e3}{f5}
\ncline[linewidth=1.0pt]{e3}{f6}
\ncline[linewidth=1.0pt]{e4}{f7}
\ncline[linewidth=1.0pt]{e4}{f8}
\psline[linewidth=1.0pt,linestyle=dotted](-2.0,-7.0)(2.0,-7.0)
\end{center}
\vskip10truecm
\caption{Cayley tree of $h$ generations, labelled from $g=1$ to $g=h$, with generation $g=1$ being the central chain (in bold at the top of the tree) of $r$ sites used for calculation of the correlation function. In the infinite limit, the deep interior of this form of the Cayley tree has the same properties as the more standard form, where there is a single central (root) site, rather than a chain. \label{fig:corrCayleyM}}
\end{figure}

\section{Spin-spin Correlation Function}
\label{sec:CorrFunc}

The spin-spin correlation function, $C(r)$, between two spins labelled $i=0$ and $i=r$, separated by the shortest chemical distance $r>0$, is given by, 
\begin{eqnarray}
C(r)=\left\langle s_0s_r\right\rangle-\left\langle s_0\right\rangle\left\langle s_r\right\rangle~,
\label{eq:CorrFunc}
\end{eqnarray}
where $\langle ...\rangle$ denotes an average over quenched disorder. 
In order to evaluate the correlation function, it is convenient to define a Cayley tree of height $h$ by dividing it into a set of $h$ generations $g$ of spins (see \fref{fig:corrCayleyM}). 
The first generation, $g=1$, forms a central chain consisting of $r+1$ connected spins, in contrast to the standard definition with a single central (root) site (presented in~\ref{sec:Dhar}).
The two boundary spins of the central chain, $i=0$ and $i=r$, are each connected to one spin within the central chain, and to $q-1$ spins in generation $g=2$ of the Cayley tree.
The interior spins of the central chain $0<i<r$ interact with $2$ other neighbouring spins (see thick horizontal lines in \fref{fig:corrCayleyM}) in the central chain and $q-2$ second generation spins. 
Each spin in generation $g>1$ then interacts with a single spin in generation $g-1$, and $q-1$
spins in generation $g+1$.
Spins in the last (boundary) generation, $g=h$, interact only with a single spin in generation $h-1$.
In the limit of large number of generations, a Cayley tree defined in this way tends to the same limit, i.e. a Bethe lattice, as the more standard Cayley tree defined in~\cite{Dhar1997,Baxter_Book}.

For finite values of external magnetic field, spins in all generations are relaxed.
First, we relax the spins in generations $g>1$, so that the
spins in generation $g=2$ are up with probability $P^*$, which is the solution to the self-consistent equation,
\begin{eqnarray}
P^*=F(P^*)~.
\label{eq:selfConsistantM}
\end{eqnarray}
The function $F(P^*)$ is given by \cite{Dhar1997},
\begin{eqnarray}
F(P^*)=\sum_{m=0}^{q-1}{{q-1}\choose m}{\left[P^*\right]}^m{\left[1-P^*\right]}^{q-1-m}\nonumber\\ 
I(-H+(q-2m)J,\infty)~,
\label{eq:Fm}
\end{eqnarray}
with,
\begin{equation}
I(h',h'')=\int\limits_{h'}^{h''}\rho(h_i)\,\rm{d}h_i~.
\label{eq:integral}
\end{equation}
Then, knowing the state of spins in generation $g=2$, we relax the spins in the central chain $g=1$ and calculate the correlation function (see Appendix A).

The relaxed central chain can contain both flipped and unflipped spins. 
In order to know the state of spin $i\in [0,r]$ we need to know the local field, $f_i$, which depends on the neighbourhood of this spin and the random field, $h_i$,
\begin{eqnarray}
f_i\left(h_i,n_i^\prime,n_i^{\prime\prime}\right)=
H+h_i - (2(n_i^\prime+n_i^{\prime\prime})-q)J~.
\label{eq:local_field}
\end{eqnarray}
Here, the state of the neighbourhood is represented by the number $n_i^\prime$ of flipped neighbours in the central chain and the number $n_i^{\prime\prime}$ of flipped neighbours in generation $2$ prior to the relaxation of the central chain, so that the total number of flipped neighbours $n_i=n^\prime_i+n^{\prime\prime}_i$.

According to the values of the local fields,
$f_i\left(h_i,n_i^\prime,n_i^{\prime\prime}\right)$, where the
variables  $h_i$ and $n_i^{\prime\prime}$ are fixed during the
relaxation of the central chain, spins in nodes $0\leq i\leq r$ can be divided
into three categories: 
($1$) those that experience a positive local field when all neighbours within the central
chain are down $n^\prime_i=0$,
i.e. spins in the set $R_1=\{(n_i^{\prime\prime},h_i)\,|\,f_i\left(h_i,0,n_i^{\prime\prime}\right)>0\}$;
($2$) those that only experience a positive local field after one of their two neighbours in the central chain has flipped
i.e. $R_2=\{(n_i^{\prime\prime},h_i)\,|\,(f_i\left(h_i,0,n_i^{\prime\prime}\right)\leq
0 \rm{~and~}
f_i\left(h_i,1,n_i^{\prime\prime}\right)>0)\}$; 
($3$) all other spins
$R_3=\{(n_i^{\prime\prime},h_i)\,|\,f_i\left(h_i,1,n_i^{\prime\prime}\right)\leq
0\}$.

In general, the states of the chain-boundary spins $i=0$ and $i=r$ are determined entirely by the categories $c_i$ of all the spins in the central chain.
Therefore, one can calculate the value of $\langle s_0s_r\rangle$ given the probabilities $Q_c\equiv \widehat{Q}_c(q-2) $ and $Q_c^\prime \equiv \widehat{Q}_c(q-1)$ of interior and boundary spins in the central chain falling into category $c=1,2,$ or $3$, where,
\begin{eqnarray}
&~&\widehat{Q}_c(\tilde{q})=
\sum_{n^{\prime\prime}=0}^{\tilde{q}}P(n^{\prime\prime})P\left((n^{\prime\prime},h_i)\in R_c|n^{\prime\prime}\right)\nonumber\\
&=&
\sum_{n^{\prime\prime}=0}^{\tilde{q}}{{\tilde{q}}\choose{n^{\prime\prime}}}
\left[P^*\right]^{n^{\prime\prime}}\left[1-P^*\right]^{{\tilde{q}}-n^{\prime\prime}}
I(h_{\rm{min}},h_{\rm{max}})~,
\label{eq:Q}
\end{eqnarray}
and $h_{\rm{min}}$ and $h_{\rm{max}}$ are given by,
\begin{equation}
(h_{\rm{min}},~h_{\rm{max}})= 
\cases{ 
\left(-H-J[2n^{\prime\prime}-q],~\infty\right),& c=1\\
(-H-J[2n^{\prime\prime}+2-q],& \\
~~~~~~~~~-H-J[2n^{\prime\prime}-q]),& c=2\\
\left(-\infty,~-H-J[2n^{\prime\prime}+2-q]\right),& c=3
}
\label{eq:limits}
\end{equation}
The value of $\widehat{Q}_c(\tilde{q})$ given by \eref{eq:Q} is a sum of probabilities 
$P(n^{\prime\prime})$ that $n^{\prime\prime}$ neighbours in generation $g=2$ have flipped, multiplied by the conditional probability, $P\left((n^{\prime\prime},h_i)\in R_c|n^{\prime\prime}\right)$, that the spin is in category $c$ given this value of $n^{\prime\prime}$.
This conditional probability translates into the integral $I(h_{\rm{min}},~h_{\rm{max}})$ over the range of random fields specified by \eref{eq:limits} which cause the spin to be in that category.

Avalanches within the central chain are nucleated at all sites of category $1$, and propagate through sites of category $2$, terminating either at the end of the chain, or at a site of category $3$.
Therefore, for a given configuration of categories $c_i$ of spins in nodes $0<i<r$ it is possible to determine the state of the spins at the end of the chain, $i=0$ and $i=r$, by finding the first site from each end which is not of category $2$.
If this first site not of category $2$ is of category $1$ then the respective end spin will be up and if such first site is of category $3$ the end spin will be down (see Appendix B).
Summing over all combinations of such systems it is possible to calculate $\langle s_0s_r\rangle$.
Similarly, it is possible to consider just one spin $\langle s_0\rangle$.
Therefore the correlation function can be found by the formula~\eref{eq:CorrFunc}, and is given by,
\begin{eqnarray}
C(r)&=&\left\{\frac{4Q_2^\prime Q_1Q_3}{(1-Q_2)^2Q_2}\left[\left(\frac{Q_1^\prime} {Q_1}+\frac{Q_3^\prime}{Q_3}\right)\left(1-Q_2\right)+\right.\right.\nonumber\\
& &\left.\left. \left(2Q_2^\prime-\frac{Q_2^\prime}{Q_2}\right)\right]+
\frac{4Q_1Q_3{Q_2^\prime}^2}{(1-Q_2)Q_2^2}r\right\}\left(Q_2\right)^r~.\label{eq:corrFuncFinal}
\end{eqnarray}
\Eref{eq:corrFuncFinal} is the main result of our derivation, and represents the exact expression for the correlation function in a Bethe lattice.
For $r \gg 1$, the correlation function should follow the behaviour~\cite{Hu1997,Baxter_Book},
\begin{equation}
C(r)\sim {{1}\over{(q-1)^{r}r^{\tilde{\eta}}}}\exp\left({-r \over \xi}\right)~.
\label{eq:corrFuncHyp}
\end{equation}
where, by comparing~\eref{eq:corrFuncHyp} with~\eref{eq:corrFuncFinal} in the limit of large $r$, it is found that $\tilde{\eta} =-1$ and the correlation length is given by,
\begin{equation}
\xi=\left[-\ln \left((q-1)Q_2\right)\right]^{-1}~.
\label{eq:xi}
\end{equation}

The exact expression for correlation function given by~\eref{eq:corrFuncFinal} can be supported numerically 
by evaluation of the correlation function on random $q$-regular graphs. These are random graphs with a fixed coordination number $q$ whose topology at local scales is similar to that of a Cayley tree with the same coordination number~\cite{Marinari_JSTAT2004}. The absence of boundaries in $q$-regular graphs make them more suitable for numerical simulations than the Bethe lattice used in our analytical derivation (see Appendix C). \Fref{fig:numerics} demonstrates that the results of numerical calculations closely match the analytical formula,~\eref{eq:corrFuncFinal}.

\begin{figure}
\begin{center}
\includegraphics[clip=true,width=6cm]{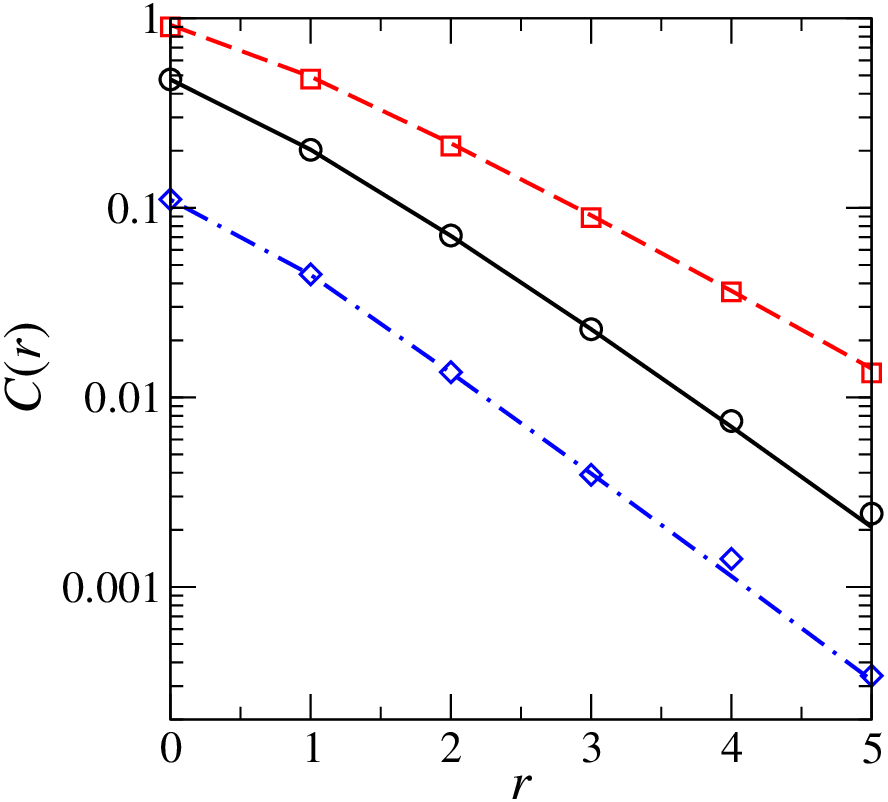}
\end{center}
\caption{The dependence of the correlation function $C(r)$ on $r$ for $q=4$ according to~\eref{eq:corrFuncFinal} (lines) and numerical simulations (symbols) for several values of external field: $H=0.75$ (circles, solid line), $1.0$ (squares, dashed line), and $1.25$ (diamonds, dot-dashed line).
All results were obtained for systems characterised by $\Delta=\Delta_c=1.78125895$.
Numerical simulations consider a $q$-regular graph ($q=4$) of $N=10^5$ spins, and calculate the mean of the correlation function between all pairs of sites in the system for $10^3$ realisations of quenched disorder.
\label{fig:numerics}}
\end{figure}

\section{Expansion Around Criticality}
\label{sec:expand}

Near the critical point the correlation length diverges according to standard scaling relations,
\begin{eqnarray}
\xi&\propto& (\Delta-\Delta_c)^{-\nu}~,~~~\rm{if}~~~H=H_c\label{eq:expnu}\\
\xi&\propto& (H-H_c)^{-\mu}~,~~~\rm{if}~~~\Delta=\Delta_c~,\label{eq:expmu}
\end{eqnarray}
with $\nu$ and $\mu$ being the critical exponents.
In order to find these exponents, it is convenient to rewrite~\eref{eq:xi} 
using the following property of $Q_2$,
\begin{eqnarray}
Q_{2}= \frac{1}{(q-1)}\, \frac{\partial F(P^*)}{\partial P^*}~.
\label{eq:dF}
\end{eqnarray}
Introducing this expression in~\eref{eq:xi}, one obtains,
\begin{eqnarray}
\xi=\left[-\ln \left(1+{\partial G\over\partial P^*}\right)\right]^{-1}~,\label{eq:xiInG}
\end{eqnarray}
where $G(P^*)=F(P^*)-P^*$.
\Eref{eq:xiInG} can be expanded in terms of $\partial G/\partial P^*$ which is equal to $0$ at criticality and small nearby, resulting in,
\begin{eqnarray}
\xi=-\left(\frac{\partial G}{\partial P^*}\right)^{-1}+\rm{O}(1)\label{eq:xiExpand}
\end{eqnarray}
In the case of Bethe lattices with coordination numbers $q=2$ or $q=3$, there is no transition at any non-zero disorder. 
However, for such coordination numbers, it is possible to show by our method that the correlation length diverges exponentially as disorder goes to zero. This result is in agreement with the calculation given in \cite{Kimball2002} for the case $q=2$.

For higher coordination numbers $q\geq 4$, where a transition exists, expansion of~\eref{eq:xiExpand} around criticality can be done in terms of $H-H_c$, $\Delta-\Delta_c$ and $P^*-P^*_c$. 
Using the known mean-field exponents $\delta=3$ and $\beta=1/2$~\cite{Sethna1993,Dhar1997} describing the behaviour of $P^*-P_c$ in terms of $H-H_c$ and $\Delta-\Delta_c$, respectively, around criticality and noting that both the first and second derivatives $\partial G/\partial P^*=\partial^2 G/\partial {P^*}^2=0$ one obtains the values $\nu=\nu_{\rm{MF}}\bar{d}=1$ and $\mu=\mu_{\rm{MF}}\bar{d}=2/3$ for the correlation length critical exponents~(see Appendix D). 

The values of these exponents can be explained in terms of the mean-field values using the argument for percolation of~\cite{Straley1982,deGennes1976}.
In this argument, it is demonstrated that the critical percolation cluster in high dimensions ($d\geq 6$) is sparse enough as for the loops to be unimportant.
In this case, the critical percolation cluster is essentially the same as that on a Bethe lattice, i.e. its backbone consists of chains of links (which form random walks), with occasional branch points (nodes).
The mean chain length between branching nodes is defined as the correlation length which scales as $(p-p_c)^{-\nu_{\rm{perc}}}$ (where $p$ and $p_c$ are bond probability and the critical bond probability) with the power $\nu_{\rm{perc}}=1$ for chemical distance and $\nu_{\rm{perc}}=1/2$ for Euclidean distance.
The same arguments can be applied to the critical avalanche clusters in the zt-RFIM on a hypercubic lattice and Bethe lattice, so that the correlation length should scale with the exponents $\mu=2/3$, $\nu=1$ for chemical distance through the avalanche cluster and $\mu=1/3$, $\nu=1/2$ for Euclidean distance across the hypercubic lattice.

\section{Conclusions}
\label{sec:Conclude}

To conclude, we derived an exact analytical formula for the spin-spin correlation function for the zt-RFIM defined on a Bethe lattice and investigated the scaling of the correlation length near criticality.
We demonstrate that if the correlation length is defined not to include the intrinsic topological length scale, then it diverges around criticality with exponents consistent with the mean-field description. 
In fact, choosing the dimensionality of the Bethe lattice to be a function of the length scale $r$~\cite{Hu1997,Baxter_Book}, that is, replacing the term $r^d$ in the standard definition of the correlation length with $(q-1)^r$, allows for analysis consistent with that performed for hypercubic lattices.

\section{Acknowledgements}

We thank M. L. Rosinberg for fruitful discusions. 
T.P.H. would like to thank the UK EPSRC for financial support. 

\appendix

\section{Relaxation of Generations $g>1$ of the Cayley tree}\label{sec:Dhar}

\begin{figure}
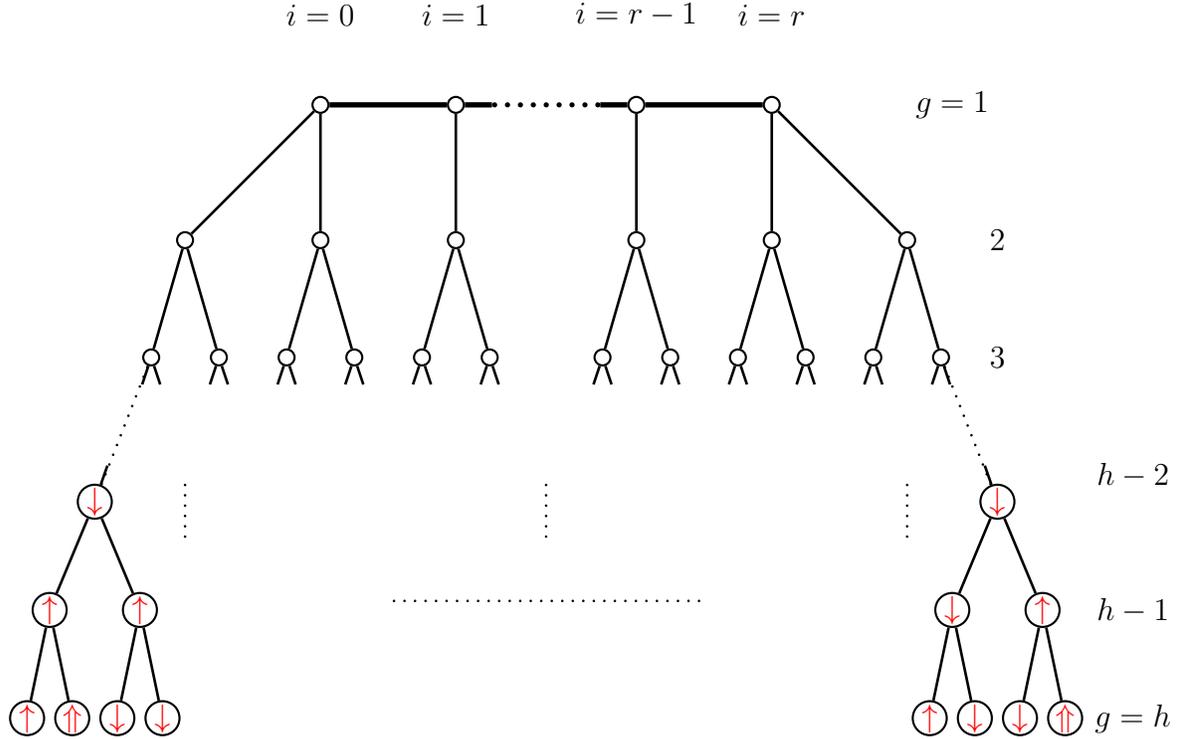

\vskip1.5truecm
\begin{center}
\psset{unit=1.2cm}
\rput(4.5,0.0){ $g=1$ }
\rput(-2.5,1.0){ $i=0$ }
\cnode[linewidth=0.8pt](-2.5,0.0){0.1}{a1}
\rput(-1.0,1.0){ $i=1$ }
\cnode[linewidth=0.8pt](-1.0,0.0){0.1}{a2}
\cnode[linewidth=0pt](-0.6,0.0){0}{a2p1}
\cnode[linewidth=0pt](0.6,0.0){0}{a2p2}
\rput(1.0,1.0){ $i=r-1$ }
\cnode[linewidth=0.8pt](1.0,0.0){0.1}{a3}
\rput(2.5,1.0){ $i=r$ }
\cnode[linewidth=0.8pt](2.5,0.0){0.1}{a4}
\ncline[linewidth=2.0pt]{a1}{a2}
\ncline[linewidth=2.0pt]{a2}{a2p1}
\ncline[linewidth=2.0pt,linestyle=dotted]{a2p1}{a2p2}
\ncline[linewidth=2.0pt]{a2p2}{a3}
\ncline[linewidth=2.0pt]{a3}{a4}
\rput(5,-1.5){ $2$ }
\cnode[linewidth=0.8pt](-4.0,-1.5){0.1}{b1}
\cnode[linewidth=0.8pt](-2.5,-1.5){0.1}{b2}
\cnode[linewidth=0.8pt](-1.0,-1.5){0.1}{b3}
\cnode[linewidth=0.8pt](1.0,-1.5){0.1}{b4}
\cnode[linewidth=0.8pt](2.5,-1.5){0.1}{b5}
\cnode[linewidth=0.8pt](4.0,-1.5){0.1}{b6}
\ncline[linewidth=1.0pt]{a1}{b1}
\ncline[linewidth=1.0pt]{a1}{b2}
\ncline[linewidth=1.0pt]{a2}{b3}
\ncline[linewidth=1.0pt]{a3}{b4}
\ncline[linewidth=1.0pt]{a4}{b5}
\ncline[linewidth=1.0pt]{a4}{b6}
\rput(5,-2.8){ $3$ }
\cnode[linewidth=0.8pt](-4.375,-2.8){0.1}{c1}
\cnode[linewidth=0.8pt](-3.625,-2.8){0.1}{c2}
\cnode[linewidth=0.8pt](-2.875,-2.8){0.1}{c3}
\cnode[linewidth=0.8pt](-2.125,-2.8){0.1}{c4}
\cnode[linewidth=0.8pt](-1.375,-2.8){0.1}{c5}
\cnode[linewidth=0.8pt](-0.625,-2.8){0.1}{c6}
\cnode[linewidth=0.8pt](0.625,-2.8){0.1}{c7}
\cnode[linewidth=0.8pt](1.375,-2.8){0.1}{c8}
\cnode[linewidth=0.8pt](2.125,-2.8){0.1}{c9}
\cnode[linewidth=0.8pt](2.875,-2.8){0.1}{c10}
\cnode[linewidth=0.8pt](3.625,-2.8){0.1}{c11}
\cnode[linewidth=0.8pt](4.375,-2.8){0.1}{c12}
\ncline[linewidth=1.0pt]{b1}{c1}
\ncline[linewidth=1.0pt]{b1}{c2}
\ncline[linewidth=1.0pt]{b2}{c3}
\ncline[linewidth=1.0pt]{b2}{c4}
\ncline[linewidth=1.0pt]{b3}{c5}
\ncline[linewidth=1.0pt]{b3}{c6}
\ncline[linewidth=1.0pt]{b4}{c7}
\ncline[linewidth=1.0pt]{b4}{c8}
\ncline[linewidth=1.0pt]{b5}{c9}
\ncline[linewidth=1.0pt]{b5}{c10}
\ncline[linewidth=1.0pt]{b6}{c11}
\ncline[linewidth=1.0pt]{b6}{c12}
\cnode[linewidth=0pt](-4.275,-3.1){0.0}{c1p1}
\cnode[linewidth=0pt](-3.525,-3.1){0.0}{c2p1}
\cnode[linewidth=0pt](-2.775,-3.1){0.0}{c3p1}
\cnode[linewidth=0pt](-2.025,-3.1){0.0}{c4p1}
\cnode[linewidth=0pt](-1.275,-3.1){0.0}{c5p1}
\cnode[linewidth=0pt](-0.525,-3.1){0.0}{c6p1}
\cnode[linewidth=0pt](0.525,-3.1){0.0}{c7p1}
\cnode[linewidth=0pt](1.275,-3.1){0.0}{c8p1}
\cnode[linewidth=0pt](2.025,-3.1){0.0}{c9p1}
\cnode[linewidth=0pt](2.775,-3.1){0.0}{c10p1}
\cnode[linewidth=0pt](3.525,-3.1){0.0}{c11p1}
\cnode[linewidth=0pt](4.275,-3.1){0.0}{c12p1}
\cnode[linewidth=0pt](-4.475,-3.1){0.0}{c1p2}
\cnode[linewidth=0pt](-3.725,-3.1){0.0}{c2p2}
\cnode[linewidth=0pt](-2.975,-3.1){0.0}{c3p2}
\cnode[linewidth=0pt](-2.225,-3.1){0.0}{c4p2}
\cnode[linewidth=0pt](-1.475,-3.1){0.0}{c5p2}
\cnode[linewidth=0pt](-0.725,-3.1){0.0}{c6p2}
\cnode[linewidth=0pt](0.725,-3.1){0.0}{c7p2}
\cnode[linewidth=0pt](1.475,-3.1){0.0}{c8p2}
\cnode[linewidth=0pt](2.225,-3.1){0.0}{c9p2}
\cnode[linewidth=0pt](2.975,-3.1){0.0}{c10p2}
\cnode[linewidth=0pt](3.725,-3.1){0.0}{c11p2}
\cnode[linewidth=0pt](4.475,-3.1){0.0}{c12p2}
\ncline[linewidth=1.0pt]{c1}{c1p1}
\ncline[linewidth=1.0pt]{c2}{c2p1}
\ncline[linewidth=1.0pt]{c3}{c3p1}
\ncline[linewidth=1.0pt]{c4}{c4p1}
\ncline[linewidth=1.0pt]{c5}{c5p1}
\ncline[linewidth=1.0pt]{c6}{c6p1}
\ncline[linewidth=1.0pt]{c7}{c7p1}
\ncline[linewidth=1.0pt]{c8}{c8p1}
\ncline[linewidth=1.0pt]{c9}{c9p1}
\ncline[linewidth=1.0pt]{c10}{c10p1}
\ncline[linewidth=1.0pt]{c11}{c11p1}
\ncline[linewidth=1.0pt]{c12}{c12p1}
\ncline[linewidth=1.0pt]{c1}{c1p2}
\ncline[linewidth=1.0pt]{c2}{c2p2}
\ncline[linewidth=1.0pt]{c3}{c3p2}
\ncline[linewidth=1.0pt]{c4}{c4p2}
\ncline[linewidth=1.0pt]{c5}{c5p2}
\ncline[linewidth=1.0pt]{c6}{c6p2}
\ncline[linewidth=1.0pt]{c7}{c7p2}
\ncline[linewidth=1.0pt]{c8}{c8p2}
\ncline[linewidth=1.0pt]{c9}{c9p2}
\ncline[linewidth=1.0pt]{c10}{c10p2}
\ncline[linewidth=1.0pt]{c11}{c11p2}
\ncline[linewidth=1.0pt]{c12}{c12p2}
\psline[linewidth=1.0pt,linestyle=dotted](-4.0,-4.2)(-4.0,-4.8)
\psline[linewidth=1.0pt,linestyle=dotted](0.0,-4.2)(0.0,-4.8)
\psline[linewidth=1.0pt,linestyle=dotted](4.0,-4.2)(4.0,-4.8)
\rput(6.5,-4.1){ $h-2$ }
\cnode[linewidth=0.8pt](-5.0,-4.4){0.2}{d1}
\rput(-5.0,-4.4){ ${\red \downarrow}$ }
\cnode[linewidth=0.8pt](5.0,-4.4){0.2}{d2}
\rput(5.0,-4.4){ ${\red \downarrow}$ }
\cnode[linewidth=0.0pt](-4.86,-4.0){0.0}{d1s}
\cnode[linewidth=0.0pt](4.86,-4.0){0.0}{d2s}
\ncline[linewidth=1.0pt,linestyle=dotted]{c1}{d1}
\ncline[linewidth=1.0pt,linestyle=dotted]{c12}{d2}
\ncline[linewidth=1.0pt]{d1s}{d1}
\ncline[linewidth=1.0pt]{d2s}{d2}
\ncline[linewidth=1.0pt]{c3p1}{c3p1ds1}
\ncline[linewidth=1.0pt]{c3p1ds2}{d1}
\ncline[linewidth=1.0pt]{c11p1}{c11p1ds1}
\ncline[linewidth=1.0pt]{c11p1ds2}{d2}
\rput(6.5,-5.6){ $h-1$ }
\cnode[linewidth=0.8pt](-5.5,-5.6){0.2}{e1}
\rput(-5.5,-5.6){ ${\red \uparrow}$ }
\cnode[linewidth=0.8pt](-4.5,-5.6){0.2}{e2}
\rput(-4.5,-5.6){ ${\red \uparrow}$ }
\cnode[linewidth=0.8pt](4.5,-5.6){0.2}{e3}
\rput(4.5,-5.6){ ${\red \downarrow}$ }
\cnode[linewidth=0.8pt](5.5,-5.6){0.2}{e4}
\rput(5.5,-5.6){ ${\red \uparrow}$ }
\ncline[linewidth=1.0pt]{d1}{e1}
\ncline[linewidth=1.0pt]{d1}{e2}
\ncline[linewidth=1.0pt]{d2}{e3}
\ncline[linewidth=1.0pt]{d2}{e4}
\rput(6.5,-6.8){ $g=h$ }
\cnode[linewidth=0.8pt](-5.75,-6.8){0.2}{f1}
\rput(-5.75,-6.8){ ${\red \uparrow}$ }
\cnode[linewidth=0.8pt](-5.25,-6.8){0.2}{f2}
\rput(-5.25,-6.8){ ${\red \Uparrow}$ }
\cnode[linewidth=0.8pt](-4.75,-6.8){0.2}{f3}
\rput(-4.75,-6.8){ ${\red \downarrow}$ }
\cnode[linewidth=0.8pt](-4.25,-6.8){0.2}{f4}
\rput(-4.25,-6.8){ ${\red \downarrow}$ }
\cnode[linewidth=0.8pt](4.25,-6.8){0.2}{f5}
\rput(4.25,-6.8){ ${\red \uparrow}$ }
\cnode[linewidth=0.8pt](4.75,-6.8){0.2}{f6}
\rput(4.75,-6.8){ ${\red \downarrow}$ }
\cnode[linewidth=0.8pt](5.25,-6.8){0.2}{f7}
\rput(5.25,-6.8){ ${\red \downarrow}$ }
\cnode[linewidth=0.8pt](5.75,-6.8){0.2}{f8}
\rput(5.75,-6.8){ ${\red \Uparrow}$ }
\ncline[linewidth=1.0pt]{e1}{f1}
\ncline[linewidth=1.0pt]{e1}{f2}
\ncline[linewidth=1.0pt]{e2}{f3}
\ncline[linewidth=1.0pt]{e2}{f4}
\ncline[linewidth=1.0pt]{e3}{f5}
\ncline[linewidth=1.0pt]{e3}{f6}
\ncline[linewidth=1.0pt]{e4}{f7}
\ncline[linewidth=1.0pt]{e4}{f8}
\psline[linewidth=1.0pt,linestyle=dotted](-1.7,-5.5)(1.7,-5.5)
\end{center}
\vskip8truecm
\caption{Cayley tree of height $h$, drawn with each row of spins representing a generation $g$, $1\leq g\leq h$. 
The first generation, being the top of the tree, is a central chain of length $r$.
The arrows in generations $g=h$, $g=h-1$ and $g=h-2$ represent the state of the spins after the second relaxation step, $\ell=2$.
In generation $g=h$, spins marked with $\downarrow$ remain in the initial down-state after the $\ell=2$ relaxation, spins marked with $\uparrow$ flipped on the first step of relaxation, $\ell=1$, and spins marked with $\Uparrow$ flipped on the second step of relaxation, $\ell=2$, owing to the increased local field caused by the spin flips occurring directly above them.
\label{fig:corrCayley}}
\end{figure}

In this section, we describe the relaxation of spins in all generations with $g>1$ using a procedure similar to that described by \cite{Dhar1997} for obtaining the magnetisation in the zero-temperature random field Ising model (zt-RFIM). 
The system of spins within the ferromagnetic zt-RFIM, relaxing according to zero-temperature Glauber dynamics, obeys the Abelian property~\cite{Sethna1993}, i.e. the final metastable state of the spin system after relaxation is independent of the order in which spins relax.
We therefore choose a relaxation order convenient to analysis. 
We start relaxation with boundary generation $g=h$ at the bottom of the tree (see \fref{fig:corrCayley}), 
At the first step, the spins in all generations except the boundary generation $h$ are held 
in the their initial down ($\downarrow$, $s=-1$) state, while allowing any spin in generation $h$ to flip (change state, $s \rightarrow -s$) if the local field acting on that spin is positive, and thus the flip 
reduces the overall energy (see the up spins, $\uparrow$, in \fref{fig:corrCayley} in generation $g=h$). 
At the next step, spins in generation $h-1$ are similarly allowed to flip if doing so reduces the overall energy (see the up spins, $\uparrow$, in \fref{fig:corrCayley} in generation $g=h-1$), while the sites in generations $g<h-1$ are again kept fixed down.
Spins flipping in generation $h-1$ will cause a change in the local field acting at their non-flipped neighbours in generation $h$, which may become positive and thus cause these spins to flip as well (see the up spins, $\Uparrow$, in \fref{fig:corrCayley} in generation $g=h$).
Such secondary flips in generation $h$ cannot influence the spin in the above generation, which is already flipped and thus in its final state (at zero temperature).
Because of the loopless structure of the Cayley tree, these secondary flips are isolated from the above generations. 
Therefore, they cannot affect those generations and, in particular, the central chain ($g=1$) in which we are interested for calculating the correlation function.
At subsequent steps, relaxation progresses up the generations in a similar way until the spins in generation $g=2$ are relaxed.

The above relaxation procedure is a random process, due to
the contribution of quenched independent random fields $\{h_i^{(g)} | g=1,\ldots,h\}$ to the local fields $\{f_i^{(g)}|g=1,\ldots,h\}$ at generation $g$.
At the first step, all other contributions to the local fields $\{f_i^{(h)}\}$ are identical for all sites, meaning that the local fields are independent identically distributed random variables.
As a consequence any spin $s_i$ in generation $h$ may flip independently of all others in this generation with probability $P^{(h)}$, the probability that $f_i^{(h)}>0$.
At the second step, randomness in the local fields at sites in generation $h-1$ is introduced by the random fields $\{h_i^{(h-1)}\}$ at those sites in combination with the field produced by a random number of flipped
neighbours in generation $h$, both of which are independent random variables.
Therefore, 
all the flips in generation $h-1$ at the second step are independent events, occurring with probability $P^{(h-1)}$.
Similarly, according to this relaxation procedure, 
at the $\ell$-th step, the flips in generation $h-\ell+1$ are all independent events.

In order to calculate the values of the probabilities $P^{(g)}$, consider first an arbitrary spin $s_i$ in generation $g<h$ and calculate the conditional probability $p_n$ of it flipping given that it is surrounded by $n$ spins in the up state.
The condition for the spin to flip is that the local field, 
\begin{eqnarray}
f_i^{(g)}=H+h_i^{(g)}+J\sum_{\langle i|j\rangle}s_j~,
\end{eqnarray}
is greater than zero, i.e. 
\begin{eqnarray}
f_i^{(g)}(h_i^{(g)};n_i)= H+h_i^{(g)}-J(q-2n_i^{(g)})>0~.
\label{eq:flipCondition}
\end{eqnarray}
Here, $n_i^{(g)}\equiv n_i=\sum_{\langle i|j \rangle}{(1+s_j)/2}$ gives the number of neighbouring spins $j$ (in generations $g\pm 1$) of the spin $i$ (in generation $g$) which are in the up state, $s_j=1$.
The probability of the random field $h_i$ meeting condition~\eref{eq:flipCondition} is,
\begin{equation}
p_n=\int_{-\infty}^{\infty}\Theta\left( f_i(h_i;n) \right)\rho(h_i)dh_i=I(-H+(q-2n)J,\infty)~,
\label{eq:pn}
\end{equation}
where $\Theta(x)$ stands for the Heaviside step function, and $I(h_{\rm{min}},h_{\rm{max}})$ is given by~\eref{eq:integral}.

The variable $n$ in~\eref{eq:pn} is a random variable depending on the orientation of neighbouring spins, which are located in generations $g_\pm=g\pm 1$.
For $1<g<h$, when a spin in generation $g$ is first relaxed (at relaxation step $h-g+1$), the neighbouring spin in generation $g_-$ is in the down-state, while the $q-1$ neighbouring spins in generation $g_+$ may each independently (as stated above) be in the up-state with probability $P^{(g+1)}$.
Therefore, the total number of neighbouring spins which are in the up-state is binomially distributed according to $n\sim B(q-1,P^{(g+1)})$, i.e. the probability of $n$ neighbours of a spin at generation $g$ being up is given by 
\[{F}(n,g)={{q-1}\choose n}{\left[P^{(g+1)}\right]}^n{\left[1-P^{(g+1)}\right]}^{q-1-n}~. 
\]
The probability $P^{(g)}$ can then be found as a sum over all
possible configurations of $n$ neighbouring spins in the
  up-state ($0\leq n \leq q-1$), 
\begin{equation}
P^{(g)}=\sum_{n=0}^{q-1}{F}(n,g)p_n=\sum_{n=0}^{q-1}{{q-1}\choose n}{\left[P^{(g+1)}\right]}^n{\left[1-P^{(g+1)}\right]}^{q-1-n}p_n~,
\label{eq:Pg}
\end{equation}
which is a recursion relation for $P^{(g)}$ valid for $1<g<h$.
 Far from the boundary of very large Cayley trees ($1<g \ll h$, $h \gg 1$), $P^{(g)}$ tends to some limiting value, $P^*=\lim_{h\to\infty}P^{(g)}$, identical for all interior generations.
This leads to the self-consistent equation for $P^*$ (see~\eref{eq:selfConsistantM}~and~\eref{eq:Fm}),
\begin{eqnarray}
F(P^*)&=&\sum_{m=0}^{q-1}{{q-1}\choose m}{\left[P^*\right]}^m{\left[1-P^*\right]}^{q-1-m}p_m~,\label{eq:F}
\\
P^*&=&F(P^*)~,\label{eq:selfConsistant}
\end{eqnarray}
so that $P^{(g)}=P^*$ for all $g>1$, i.e. after the relaxation of all spins except those in the central chain, spins in generation $2$ are in the up-state with probability $P^*$.

\begin{figure}
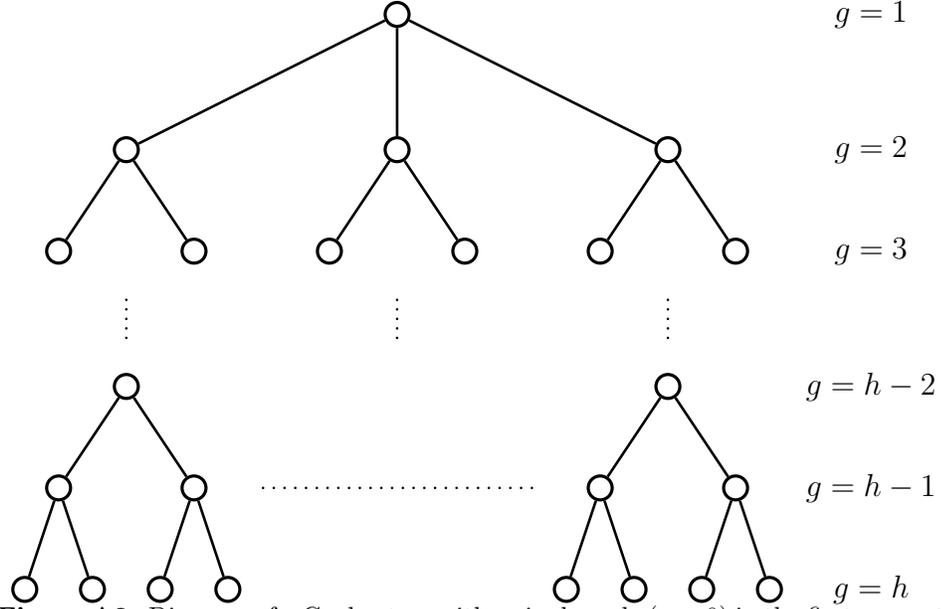

\vskip2truecm
\begin{center}
\psset{unit=0.9cm}
\rput(7.0,0.0){ $g=1$ }
\cnode[linewidth=1.2pt](0.0,0.0){0.2}{a1}
\rput(7.0,-2.0){ $g=2$ }
\cnode[linewidth=1.2pt](-4.0,-2.0){0.2}{b1}
\cnode[linewidth=1.2pt](0.0,-2.0){0.2}{b2}
\cnode[linewidth=1.2pt](4.0,-2.0){0.2}{b3}
\ncline[linewidth=1.0pt]{a1}{b1}
\ncline[linewidth=1.0pt]{a1}{b2}
\ncline[linewidth=1.0pt]{a1}{b3}
\rput(7.0,-3.5){ $g=3$ }
\cnode[linewidth=1.2pt](-5.0,-3.5){0.2}{c1}
\cnode[linewidth=1.2pt](-3.0,-3.5){0.2}{c2}
\cnode[linewidth=1.2pt](-1.0,-3.5){0.2}{c3}
\cnode[linewidth=1.2pt](1.0,-3.5){0.2}{c4}
\cnode[linewidth=1.2pt](3.0,-3.5){0.2}{c5}
\cnode[linewidth=1.2pt](5.0,-3.5){0.2}{c6}
\ncline[linewidth=1.0pt]{b1}{c1}
\ncline[linewidth=1.0pt]{b1}{c2}
\ncline[linewidth=1.0pt]{b2}{c3}
\ncline[linewidth=1.0pt]{b2}{c4}
\ncline[linewidth=1.0pt]{b3}{c5}
\ncline[linewidth=1.0pt]{b3}{c6}
\psline[linewidth=1.0pt,linestyle=dotted](-4.0,-4.2)(-4.0,-4.8)
\psline[linewidth=1.0pt,linestyle=dotted](0.0,-4.2)(0.0,-4.8)
\psline[linewidth=1.0pt,linestyle=dotted](4.0,-4.2)(4.0,-4.8)
\rput(7.0,-5.5){ $g=h-2$ }
\cnode[linewidth=1.2pt](-4.0,-5.5){0.2}{d1}
\cnode[linewidth=1.2pt](4.0,-5.5){0.2}{d2}
\rput(7.0,-7.0){ $g=h-1$ }
\cnode[linewidth=1.2pt](-5.0,-7.0){0.2}{e1}
\cnode[linewidth=1.2pt](-3.0,-7.0){0.2}{e2}
\cnode[linewidth=1.2pt](3.0,-7.0){0.2}{e3}
\cnode[linewidth=1.2pt](5.0,-7.0){0.2}{e4}
\ncline[linewidth=1.0pt]{d1}{e1}
\ncline[linewidth=1.0pt]{d1}{e2}
\ncline[linewidth=1.0pt]{d2}{e3}
\ncline[linewidth=1.0pt]{d2}{e4}
\rput(7.0,-8.5){ $g=h$ }
\cnode[linewidth=1.2pt](-5.5,-8.5){0.2}{f1}
\cnode[linewidth=1.2pt](-4.5,-8.5){0.2}{f2}
\cnode[linewidth=1.2pt](-3.5,-8.5){0.2}{f3}
\cnode[linewidth=1.2pt](-2.5,-8.5){0.2}{f4}
\cnode[linewidth=1.2pt](2.5,-8.5){0.2}{f5}
\cnode[linewidth=1.2pt](3.5,-8.5){0.2}{f6}
\cnode[linewidth=1.2pt](4.5,-8.5){0.2}{f7}
\cnode[linewidth=1.2pt](5.5,-8.5){0.2}{f8}
\ncline[linewidth=1.0pt]{e1}{f1}
\ncline[linewidth=1.0pt]{e1}{f2}
\ncline[linewidth=1.0pt]{e2}{f3}
\ncline[linewidth=1.0pt]{e2}{f4}
\ncline[linewidth=1.0pt]{e3}{f5}
\ncline[linewidth=1.0pt]{e3}{f6}
\ncline[linewidth=1.0pt]{e4}{f7}
\ncline[linewidth=1.0pt]{e4}{f8}
\psline[linewidth=1.0pt,linestyle=dotted](-2.0,-7.0)(2.0,-7.0)

\end{center}
\vskip7truecm
\caption{Diagram of a Cayley tree with a single node ($r=0$) in the first generation, $g=1$, as considered by~\cite{Dhar1997}.
\label{fig:CayleyStandard}}
\end{figure}

\begin{figure}
\begin{center}
\includegraphics[clip=true,width=6cm]{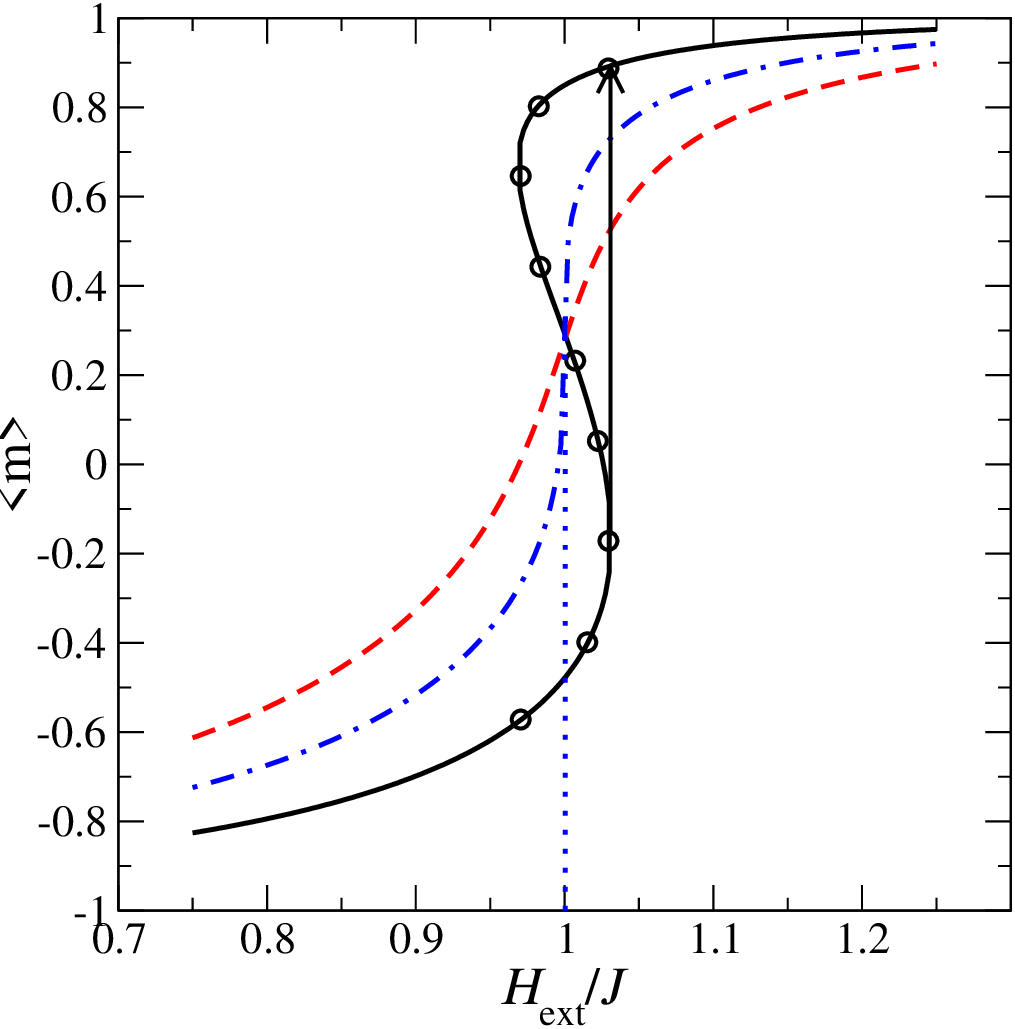}
\end{center}
\caption{\label{fig:hysteresis} Mean magnetisation, $\langle m\rangle$, vs external field, $H$, scaled by interaction strength, $J$, for zt-RFIM on a Bethe lattice of coordination number $q=4$, as magnetic field is swept upwards from $-\infty$ to $+\infty$. 
The solid, dot-dashed and dashed lines correspond to different degrees of disorder around critically: $\Delta=1.65$, $\Delta=\Delta_c=1.78215895$ and $\Delta=1.9$, respectively. The critical field ($H_c=1.0$) is marked by the dotted vertical line.
In the region where the magnetisation curve is multi-valued (marked by circles), the system follows the lower branch, so that a jump in magnetisation (represented by the vertical arrow) occurs when the external field passes the coercive field strength.}
\end{figure}

In the case of the standard Cayley tree (see \fref{fig:CayleyStandard}), the relaxation procedure described above allows the magnetisation of the central spin (generation $g=1$ in \fref{fig:CayleyStandard}) to be calculated in the following way.
As mentioned above,~\eref{eq:F} and~\eref{eq:selfConsistant} are only valid for $1<g<h$, and the central spin at generation $g=1$ should be treated separately. 
The central spin has $q$ neighbours in generation $g=2$, which, on the relaxation of that spin, are in the up-state with probability $P^{(2)}=P^*$.
The number of neighbours of the central site that are in the up-state is binomially distributed, $n\sim B(q,P^*)$ and the expression for $P^{(1)}$ is given by~\eref{eq:Pg} with $q-1$ replaced by $q$ and $P^{(g+1)}$ by $P^*$, i.e. 
\begin{equation}
P^{(1)}=\sum_{n=0}^{q}{{q}\choose n}{\left[P^*\right]}^n{\left[1-P^*\right]}^{q-n}p_n~.
\label{eq:PCentral}
\end{equation}
The value of $P^{(1)}$ calculated by solving the self-consistent equations~\eref{eq:F} and~\eref{eq:selfConsistant} and substituting into~\eref{eq:PCentral} allows the mean magnetisation of the lattice to be evaluated, $\left\langle m\right\rangle =2P^{(1)}-1$.
The result of this calculation for magnetisation vs external field is given in \fref{fig:hysteresis}, showing the known spinodal transition at low disorder~\cite{Dhar1997}. 

\section{Derivation of Correlation Function From Categories}

\begin{figure}
\vskip2truecm
\begin{center}
\psset{unit=1.0cm}
\rput(-7.0,0.0){ (a) }
\cnode[linewidth=1.2pt](-5.0,0){0.3}{sa0}
\cnode[linewidth=1.2pt](-4.0,0){0.3}{sa1}
\cnode[linewidth=1.2pt](-3.0,0){0.3}{sa2}
\cnode[linewidth=1.2pt](-2.0,0){0.3}{sa3}
\cnode[linewidth=1.2pt](-1.0,0){0.3}{sa4}
\cnode[linewidth=1.2pt](0.0,0){0.3}{sa5}
\cnode[linewidth=1.2pt](1.0,0){0.3}{sa6}
\cnode[linewidth=1.2pt](2.0,0){0.3}{sa7}
\cnode[linewidth=1.2pt](3.0,0){0.3}{sa8}
\cnode[linewidth=1.2pt](4.0,0){0.3}{sa9}
\cnode[linewidth=1.2pt](5.0,0){0.3}{sa10}
\cnode[linewidth=0pt](-5.3,-0.4){0.0}{sa0ll}
\cnode[linewidth=0pt](-5.0,-0.4){0.0}{sa0l}
\cnode[linewidth=0pt](-4.0,-0.4){0.0}{sa1l}
\cnode[linewidth=0pt](-3.0,-0.4){0.0}{sa2l}
\cnode[linewidth=0pt](-2.0,-0.4){0.0}{sa3l}
\cnode[linewidth=0pt](-1.0,-0.4){0.0}{sa4l}
\cnode[linewidth=0pt](0.0,-0.4){0.0}{sa5l}
\cnode[linewidth=0pt](1.0,-0.4){0.0}{sa6l}
\cnode[linewidth=0pt](2.0,-0.4){0.0}{sa7l}
\cnode[linewidth=0pt](3.0,-0.4){0.0}{sa8l}
\cnode[linewidth=0pt](4.0,-0.4){0.0}{sa9l}
\cnode[linewidth=0pt](5.0,-0.4){0.0}{sa10l}
\cnode[linewidth=0pt](5.3,-0.4){0.0}{sa10lr}
\ncline[linewidth=1.0pt]{sa0}{sa1}
\ncline[linewidth=1.0pt]{sa1}{sa2}
\ncline[linewidth=1.0pt]{sa2}{sa3}
\ncline[linewidth=1.0pt]{sa3}{sa4}
\ncline[linewidth=1.0pt]{sa4}{sa5}
\ncline[linewidth=1.0pt]{sa5}{sa6}
\ncline[linewidth=1.0pt]{sa6}{sa7}
\ncline[linewidth=1.0pt]{sa7}{sa8}
\ncline[linewidth=1.0pt]{sa8}{sa9}
\ncline[linewidth=1.0pt]{sa9}{sa10}
\ncline[linewidth=1.0pt]{sa0}{sa0ll}
\ncline[linewidth=1.0pt]{sa0}{sa0l}
\ncline[linewidth=1.0pt]{sa1}{sa1l}
\ncline[linewidth=1.0pt]{sa2}{sa2l}
\ncline[linewidth=1.0pt]{sa3}{sa3l}
\ncline[linewidth=1.0pt]{sa4}{sa4l}
\ncline[linewidth=1.0pt]{sa5}{sa5l}
\ncline[linewidth=1.0pt]{sa6}{sa6l}
\ncline[linewidth=1.0pt]{sa7}{sa7l}
\ncline[linewidth=1.0pt]{sa8}{sa8l}
\ncline[linewidth=1.0pt]{sa9}{sa9l}
\ncline[linewidth=1.0pt]{sa10}{sa10l}
\ncline[linewidth=1.0pt]{sa10}{sa10lr}
\rput(-5.0,-1.0){ $c_0=\alpha$ }
\rput(5.0,-1.0){ $c_r=\beta$ }
\rput(7.3,-0.0){ $k=0$, $l=0$ }
\rput(-7.0,-2.5){ (b/d) }
\cnode[linewidth=1.2pt](-5.0,-2.5){0.3}{sb0}
\cnode[linewidth=1.2pt](-4.0,-2.5){0.3}{sb1}
\cnode[linewidth=1.2pt](-3.0,-2.5){0.3}{sb2}
\cnode[linewidth=1.2pt](-2.0,-2.5){0.3}{sb3}
\cnode[linewidth=1.2pt](-1.0,-2.5){0.3}{sb4}
\cnode[linewidth=1.2pt](0.0,-2.5){0.3}{sb5}
\cnode[linewidth=1.2pt,fillstyle=solid,fillcolor=gray](1.0,-2.5){0.3}{sb6}
\cnode[linewidth=1.2pt,fillstyle=solid,fillcolor=gray](2.0,-2.5){0.3}{sb7}
\cnode[linewidth=1.2pt,fillstyle=solid,fillcolor=gray](3.0,-2.5){0.3}{sb8}
\cnode[linewidth=1.2pt,fillstyle=solid,fillcolor=gray](4.0,-2.5){0.3}{sb9}
\cnode[linewidth=1.2pt,fillstyle=solid,fillcolor=gray](5.0,-2.5){0.3}{sb10}
\cnode[linewidth=0pt](-5.3,-2.9){0.0}{sb0ll}
\cnode[linewidth=0pt](-5.0,-2.9){0.0}{sb0l}
\cnode[linewidth=0pt](-4.0,-2.9){0.0}{sb1l}
\cnode[linewidth=0pt](-3.0,-2.9){0.0}{sb2l}
\cnode[linewidth=0pt](-2.0,-2.9){0.0}{sb3l}
\cnode[linewidth=0pt](-1.0,-2.9){0.0}{sb4l}
\cnode[linewidth=0pt](0.0,-2.9){0.0}{sb5l}
\cnode[linewidth=0pt](1.0,-2.9){0.0}{sb6l}
\cnode[linewidth=0pt](2.0,-2.9){0.0}{sb7l}
\cnode[linewidth=0pt](3.0,-2.9){0.0}{sb8l}
\cnode[linewidth=0pt](4.0,-2.9){0.0}{sb9l}
\cnode[linewidth=0pt](5.0,-2.9){0.0}{sb10l}
\cnode[linewidth=0pt](5.3,-2.9){0.0}{sb10lr}
\ncline[linewidth=1.0pt]{sb0}{sb1}
\ncline[linewidth=1.0pt]{sb1}{sb2}
\ncline[linewidth=1.0pt]{sb2}{sb3}
\ncline[linewidth=1.0pt]{sb3}{sb4}
\ncline[linewidth=1.0pt]{sb4}{sb5}
\ncline[linewidth=1.0pt]{sb5}{sb6}
\ncline[linewidth=1.0pt]{sb6}{sb7}
\ncline[linewidth=1.0pt]{sb7}{sb8}
\ncline[linewidth=1.0pt]{sb8}{sb9}
\ncline[linewidth=1.0pt]{sb9}{sb10}
\ncline[linewidth=1.0pt]{sb0}{sb0ll}
\ncline[linewidth=1.0pt]{sb0}{sb0l}
\ncline[linewidth=1.0pt]{sb1}{sb1l}
\ncline[linewidth=1.0pt]{sb2}{sb2l}
\ncline[linewidth=1.0pt]{sb3}{sb3l}
\ncline[linewidth=1.0pt]{sb4}{sb4l}
\ncline[linewidth=1.0pt]{sb5}{sb5l}
\ncline[linewidth=1.0pt]{sb6}{sb6l}
\ncline[linewidth=1.0pt]{sb7}{sb7l}
\ncline[linewidth=1.0pt]{sb8}{sb8l}
\ncline[linewidth=1.0pt]{sb9}{sb9l}
\ncline[linewidth=1.0pt]{sb10}{sb10l}
\ncline[linewidth=1.0pt]{sb10}{sb10lr}
\rput(-5.0,-3.5){ $c_0=\alpha$ }
\rput(-4.0,-3.5){  }
\rput(-3.0,-3.5){  }
\rput(-2.0,-3.5){  }
\rput(-1.0,-3.5){  }
\rput(0.0,-3.5){ $\beta$ }
\rput(1.0,-3.5){ $2$ }
\rput(2.0,-3.5){ $2$ }
\rput(3.0,-3.5){ $2$ }
\rput(4.0,-3.5){ $2$ }
\rput(5.0,-3.5){ $c_r=2$ }
\uput{0}[ur](0.9,-4.3){$\underbrace{\hspace{4.6cm}}_{\normalsize l=5}$} 
\rput(7.3,-2.5){ $k=0$, $0<l<r$ }
\rput(-7.0,-5.0){ (c/e) }
\cnode[linewidth=1.2pt](-5.0,-5.0){0.3}{sc0}
\cnode[linewidth=1.2pt,fillstyle=solid,fillcolor=gray](-4.0,-5.0){0.3}{sc1}
\cnode[linewidth=1.2pt,fillstyle=solid,fillcolor=gray](-3.0,-5.0){0.3}{sc2}
\cnode[linewidth=1.2pt,fillstyle=solid,fillcolor=gray](-2.0,-5.0){0.3}{sc3}
\cnode[linewidth=1.2pt,fillstyle=solid,fillcolor=gray](-1.0,-5.0){0.3}{sc4}
\cnode[linewidth=1.2pt,fillstyle=solid,fillcolor=gray](0.0,-5.0){0.3}{sc5}
\cnode[linewidth=1.2pt,fillstyle=solid,fillcolor=gray](1.0,-5.0){0.3}{sc6}
\cnode[linewidth=1.2pt,fillstyle=solid,fillcolor=gray](2.0,-5.0){0.3}{sc7}
\cnode[linewidth=1.2pt,fillstyle=solid,fillcolor=gray](3.0,-5.0){0.3}{sc8}
\cnode[linewidth=1.2pt,fillstyle=solid,fillcolor=gray](4.0,-5.0){0.3}{sc9}
\cnode[linewidth=1.2pt,fillstyle=solid,fillcolor=gray](5.0,-5.0){0.3}{sc10}
\cnode[linewidth=0pt](-5.3,-5.4){0.0}{sc0ll}
\cnode[linewidth=0pt](-5.0,-5.4){0.0}{sc0l}
\cnode[linewidth=0pt](-4.0,-5.4){0.0}{sc1l}
\cnode[linewidth=0pt](-3.0,-5.4){0.0}{sc2l}
\cnode[linewidth=0pt](-2.0,-5.4){0.0}{sc3l}
\cnode[linewidth=0pt](-1.0,-5.4){0.0}{sc4l}
\cnode[linewidth=0pt](0.0,-5.4){0.0}{sc5l}
\cnode[linewidth=0pt](1.0,-5.4){0.0}{sc6l}
\cnode[linewidth=0pt](2.0,-5.4){0.0}{sc7l}
\cnode[linewidth=0pt](3.0,-5.4){0.0}{sc8l}
\cnode[linewidth=0pt](4.0,-5.4){0.0}{sc9l}
\cnode[linewidth=0pt](5.0,-5.4){0.0}{sc10l}
\cnode[linewidth=0pt](5.3,-5.4){0.0}{sc10lr}
\ncline[linewidth=1.0pt]{sc0}{sc1}
\ncline[linewidth=1.0pt]{sc1}{sc2}
\ncline[linewidth=1.0pt]{sc2}{sc3}
\ncline[linewidth=1.0pt]{sc3}{sc4}
\ncline[linewidth=1.0pt]{sc4}{sc5}
\ncline[linewidth=1.0pt]{sc5}{sc6}
\ncline[linewidth=1.0pt]{sc6}{sc7}
\ncline[linewidth=1.0pt]{sc7}{sc8}
\ncline[linewidth=1.0pt]{sc8}{sc9}
\ncline[linewidth=1.0pt]{sc9}{sc10}
\ncline[linewidth=1.0pt]{sc0}{sc0ll}
\ncline[linewidth=1.0pt]{sc0}{sc0l}
\ncline[linewidth=1.0pt]{sc1}{sc1l}
\ncline[linewidth=1.0pt]{sc2}{sc2l}
\ncline[linewidth=1.0pt]{sc3}{sc3l}
\ncline[linewidth=1.0pt]{sc4}{sc4l}
\ncline[linewidth=1.0pt]{sc5}{sc5l}
\ncline[linewidth=1.0pt]{sc6}{sc6l}
\ncline[linewidth=1.0pt]{sc7}{sc7l}
\ncline[linewidth=1.0pt]{sc8}{sc8l}
\ncline[linewidth=1.0pt]{sc9}{sc9l}
\ncline[linewidth=1.0pt]{sc10}{sc10l}
\ncline[linewidth=1.0pt]{sc10}{sc10lr}
\rput(-5.0,-6.0){ $c_0=\alpha$ }
\rput(-5.0,-6.5){ $=\beta$ }
\rput(-4.0,-6.0){ $2$ }
\rput(-3.0,-6.0){ $2$ }
\rput(-2.0,-6.0){ $2$ }
\rput(-1.0,-6.0){ $2$ }
\rput(0.0,-6.0){ $2$ }
\rput(1.0,-6.0){ $2$ }
\rput(2.0,-6.0){ $2$ }
\rput(3.0,-6.0){ $2$ }
\rput(4.0,-6.0){ $2$ }
\rput(5.0,-6.0){ $c_r=2$ }
\uput{0}[ur](-4.1,-6.8){$\underbrace{\hspace{9.6cm}}_{\normalsize l=r}$} 
\rput(7.3,-5.0){ $k=0$, $l=r$ }
\rput(-7.0,-7.5){ (f) }
\cnode[linewidth=1.2pt,fillstyle=solid,fillcolor=gray](-5.0,-7.5){0.3}{sf0}
\cnode[linewidth=1.2pt,fillstyle=solid,fillcolor=gray](-4.0,-7.5){0.3}{sf1}
\cnode[linewidth=1.2pt,fillstyle=solid,fillcolor=gray](-3.0,-7.5){0.3}{sf2}
\cnode[linewidth=1.2pt](-2.0,-7.5){0.3}{sf3}
\cnode[linewidth=1.2pt](-1.0,-7.5){0.3}{sf4}
\cnode[linewidth=1.2pt](0.0,-7.5){0.3}{sf5}
\cnode[linewidth=1.2pt](1.0,-7.5){0.3}{sf6}
\cnode[linewidth=1.2pt,fillstyle=solid,fillcolor=gray](2.0,-7.5){0.3}{sf7}
\cnode[linewidth=1.2pt,fillstyle=solid,fillcolor=gray](3.0,-7.5){0.3}{sf8}
\cnode[linewidth=1.2pt,fillstyle=solid,fillcolor=gray](4.0,-7.5){0.3}{sf9}
\cnode[linewidth=1.2pt,fillstyle=solid,fillcolor=gray](5.0,-7.5){0.3}{sf10}
\cnode[linewidth=0pt](-5.3,-7.9){0.0}{sf0ll}
\cnode[linewidth=0pt](-5.0,-7.9){0.0}{sf0l}
\cnode[linewidth=0pt](-4.0,-7.9){0.0}{sf1l}
\cnode[linewidth=0pt](-3.0,-7.9){0.0}{sf2l}
\cnode[linewidth=0pt](-2.0,-7.9){0.0}{sf3l}
\cnode[linewidth=0pt](-1.0,-7.9){0.0}{sf4l}
\cnode[linewidth=0pt](0.0,-7.9){0.0}{sf5l}
\cnode[linewidth=0pt](1.0,-7.9){0.0}{sf6l}
\cnode[linewidth=0pt](2.0,-7.9){0.0}{sf7l}
\cnode[linewidth=0pt](3.0,-7.9){0.0}{sf8l}
\cnode[linewidth=0pt](4.0,-7.9){0.0}{sf9l}
\cnode[linewidth=0pt](5.0,-7.9){0.0}{sf10l}
\cnode[linewidth=0pt](5.3,-7.9){0.0}{sf10lr}
\ncline[linewidth=1.0pt]{sf0}{sf1}
\ncline[linewidth=1.0pt]{sf1}{sf2}
\ncline[linewidth=1.0pt]{sf2}{sf3}
\ncline[linewidth=1.0pt]{sf3}{sf4}
\ncline[linewidth=1.0pt]{sf4}{sf5}
\ncline[linewidth=1.0pt]{sf5}{sf6}
\ncline[linewidth=1.0pt]{sf6}{sf7}
\ncline[linewidth=1.0pt]{sf7}{sf8}
\ncline[linewidth=1.0pt]{sf8}{sf9}
\ncline[linewidth=1.0pt]{sf9}{sf10}
\ncline[linewidth=1.0pt]{sf0}{sf0ll}
\ncline[linewidth=1.0pt]{sf0}{sf0l}
\ncline[linewidth=1.0pt]{sf1}{sf1l}
\ncline[linewidth=1.0pt]{sf2}{sf2l}
\ncline[linewidth=1.0pt]{sf3}{sf3l}
\ncline[linewidth=1.0pt]{sf4}{sf4l}
\ncline[linewidth=1.0pt]{sf5}{sf5l}
\ncline[linewidth=1.0pt]{sf6}{sf6l}
\ncline[linewidth=1.0pt]{sf7}{sf7l}
\ncline[linewidth=1.0pt]{sf8}{sf8l}
\ncline[linewidth=1.0pt]{sf9}{sf9l}
\ncline[linewidth=1.0pt]{sf10}{sf10l}
\ncline[linewidth=1.0pt]{sf10}{sf10lr}
\rput(-5.0,-8.5){ $c_0=2$ }
\rput(-4.0,-8.5){ $2$ }
\rput(-3.0,-8.5){ $2$ }
\rput(-2.0,-8.5){ $\alpha$ }
\uput{0}[ur](-5.5,-9.3){$\underbrace{\hspace{2.6cm}}_{\normalsize k=3}$}
\rput(-1.0,-8.5){  }
\rput(0.0,-8.5){  }
\rput(1.0,-8.5){ $\beta$ }
\rput(2.0,-8.5){ $2$ }
\rput(3.0,-8.5){ $2$ }
\rput(4.0,-8.5){ $2$ }
\rput(5.0,-8.5){ $c_r=2$ }
\rput(7.3,-7.5){ $k>0$, $l>0$ }
\rput(7.3,-7.9){ $l+k<r$ }
\uput{0}[ur](1.9,-9.3){$\underbrace{\hspace{3.6cm}}_{\normalsize l=4}$}
\rput(-7.0,-10.0){ (g) }
\cnode[linewidth=1.2pt,fillstyle=solid,fillcolor=gray](-5.0,-10.0){0.3}{sg0}
\cnode[linewidth=1.2pt,fillstyle=solid,fillcolor=gray](-4.0,-10.0){0.3}{sg1}
\cnode[linewidth=1.2pt,fillstyle=solid,fillcolor=gray](-3.0,-10.0){0.3}{sg2}
\cnode[linewidth=1.2pt,fillstyle=solid,fillcolor=gray](-2.0,-10.0){0.3}{sg3}
\cnode[linewidth=1.2pt](-1.0,-10.0){0.3}{sg4}
\cnode[linewidth=1.2pt,fillstyle=solid,fillcolor=gray](0.0,-10.0){0.3}{sg5}
\cnode[linewidth=1.2pt,fillstyle=solid,fillcolor=gray](1.0,-10.0){0.3}{sg6}
\cnode[linewidth=1.2pt,fillstyle=solid,fillcolor=gray](2.0,-10.0){0.3}{sg7}
\cnode[linewidth=1.2pt,fillstyle=solid,fillcolor=gray](3.0,-10.0){0.3}{sg8}
\cnode[linewidth=1.2pt,fillstyle=solid,fillcolor=gray](4.0,-10.0){0.3}{sg9}
\cnode[linewidth=1.2pt,fillstyle=solid,fillcolor=gray](5.0,-10.0){0.3}{sg10}
\cnode[linewidth=0pt](-5.3,-10.4){0.0}{sg0ll}
\cnode[linewidth=0pt](-5.0,-10.4){0.0}{sg0l}
\cnode[linewidth=0pt](-4.0,-10.4){0.0}{sg1l}
\cnode[linewidth=0pt](-3.0,-10.4){0.0}{sg2l}
\cnode[linewidth=0pt](-2.0,-10.4){0.0}{sg3l}
\cnode[linewidth=0pt](-1.0,-10.4){0.0}{sg4l}
\cnode[linewidth=0pt](0.0,-10.4){0.0}{sg5l}
\cnode[linewidth=0pt](1.0,-10.4){0.0}{sg6l}
\cnode[linewidth=0pt](2.0,-10.4){0.0}{sg7l}
\cnode[linewidth=0pt](3.0,-10.4){0.0}{sg8l}
\cnode[linewidth=0pt](4.0,-10.4){0.0}{sg9l}
\cnode[linewidth=0pt](5.0,-10.4){0.0}{sg10l}
\cnode[linewidth=0pt](5.3,-10.4){0.0}{sg10lr}
\ncline[linewidth=1.0pt]{sg0}{sg1}
\ncline[linewidth=1.0pt]{sg1}{sg2}
\ncline[linewidth=1.0pt]{sg2}{sg3}
\ncline[linewidth=1.0pt]{sg3}{sg4}
\ncline[linewidth=1.0pt]{sg4}{sg5}
\ncline[linewidth=1.0pt]{sg5}{sg6}
\ncline[linewidth=1.0pt]{sg6}{sg7}
\ncline[linewidth=1.0pt]{sg7}{sg8}
\ncline[linewidth=1.0pt]{sg8}{sg9}
\ncline[linewidth=1.0pt]{sg9}{sg10}
\ncline[linewidth=1.0pt]{sg0}{sg0ll}
\ncline[linewidth=1.0pt]{sg0}{sg0l}
\ncline[linewidth=1.0pt]{sg1}{sg1l}
\ncline[linewidth=1.0pt]{sg2}{sg2l}
\ncline[linewidth=1.0pt]{sg3}{sg3l}
\ncline[linewidth=1.0pt]{sg4}{sg4l}
\ncline[linewidth=1.0pt]{sg5}{sg5l}
\ncline[linewidth=1.0pt]{sg6}{sg6l}
\ncline[linewidth=1.0pt]{sg7}{sg7l}
\ncline[linewidth=1.0pt]{sg8}{sg8l}
\ncline[linewidth=1.0pt]{sg9}{sg9l}
\ncline[linewidth=1.0pt]{sg10}{sg10l}
\ncline[linewidth=1.0pt]{sg10}{sg10lr}
\rput(-5.0,-11.0){ $c_0=2$ }
\rput(-4.0,-11.0){ $2$}
\rput(-3.0,-11.0){ $2$ }
\rput(-2.0,-11.0){ $2$ }
\uput{0}[ur](-5.5,-11.8){$\underbrace{\hspace{3.6cm}}_{\normalsize k=4}$}
\rput(-1.0,-11.0){ $\alpha=\beta$ }
\rput(0.0,-11.0){ $2$ }
\rput(1.0,-11.0){ $2$ }
\rput(2.0,-11.0){ $2$ }
\rput(3.0,-11.0){ $2$ }
\rput(4.0,-11.0){ $2$ }
\rput(5.0,-11.0){ $c_r=2$ }
\uput{0}[ur](-0.1,-11.8){$\underbrace{\hspace{5.6cm}}_{\normalsize l=6}$}
\rput(7.3,-10){ $k>0$, $l>0$ }
\rput(7.3,-10.4){ $l+k=r$ }
\end{center}
\vskip11truecm
\caption{Distinct configurations of spin categories in the central chain, as used in~\eref{eq:Pkalb}.
\label{fig:CentralChain}}
\end{figure}

In order to calculate the correlation function, we need to know the states of the chain-boundary spins $i=0$ and $i=r$ which are determined entirely by the categories
$c_i$ of all the spins in the central chain.
However if the categories of the boundary spins are either $1$ or $3$ (see \sref{sec:CorrFunc} for definition),
their state depends only on their own category.
In particular, if the category of a boundary spin is $1$ then it will have a positive local field and flip.
Conversely, if a boundary spin has a category $3$ then it will have a negative local field and will never flip.
In the case that a boundary spin is of category $2$, its state depends on the categories of the other spins in the central chain.
Specifically, a boundary spin ($i=0$ or $i=r$) of category $2$ will flip only if its neighbour in the chain (either spin $i=1$ or $i=r-1$) flips first.
The state of the spins $i=1$ and $i=r-1$, given that their respective neighbours at the boundary are of category $2$, can be found in the same way.
Therefore, the state of spin $i=0$ depends on and coincides with the state of the first spin, $k$ ($0\leq k\leq r$), counting from the $i=0$ boundary, that is not of category $2$.
The state of spin $k$ is, itself, determined only by its category $\alpha=c_k$, with $c_k={1}$ or $c_k={3}$ but, by definition, $\alpha \neq 2$.
Similarly, the state of spin $r$ is determined by the state of spin $r-l$ ($0 \leq l \leq r-k$), the first spin counting from the $i=r$ boundary which is not of category $2$, being instead of category $\beta=c_{r-l}$. 
In the special case of all the spins in the central chain being of category $2$, the values of $\alpha$, $\beta$, $k$ and $l$ cannot be defined.
However, this situation is simpler because all spins in the central chain after relaxation are necessarily in the down-state, $s_i=-1$.

The probability, $P(k,\alpha;l,\beta)$, that all spins $i$ ($0\leq i<k$) are
of category $2$ with spin $k$ being of category $\alpha$, and
simultaneously all spins $j$ ($r-l<j\leq r$) are of category $2$ with spin
$r-l$ being of category $\beta$, can be calculated explicitly (see \fref{fig:CentralChain} for details).
In \fref{fig:CentralChain} we show the central chain of length $r$ with white circles representing spin of either category $1$ or $3$, and grey circles are used for spins of category $2$. 
Circles with crosses represent spins for which the category is irrelevant to the state of the boundary spins.
All distinct configurations of spin categories (excluding the special case when all spins are of category $2$) are represented by rows (a)-(g). 
In configuration (a), the boundary spin are not of category $2$, so that the categories of all other spins are irrelevant.
In configurations (b) and (d), one of the boundary spins is of category $2$, and the first spin which is not of category $2$ linked to this spin does not coincide with the other boundary spin.
In configurations (c) and (e), all the spins are of category $2$, except one of the boundary spins, meaning $\alpha=\beta$.
In configurations (f)/(g), both boundary spins are of category $2$, and the first spins not of category $2$, counting from spin $i=0$ and from $i=r$ do not coincide/do coincide.
There are two important observations to be taken into account in order to  evaluate $P(k,\alpha;l,\beta)$. 
 The first observation is that $\alpha$ and $\beta$ are independent variables only if $k+l<r$ (see configurations (a), (b), (d) and (f) in \fref{fig:CentralChain}), while otherwise, i.e. when $k+l=r$, they are equal, $\alpha=\beta$ (see configurations (c), (e) and (g)).
The second point to consider is that the configurations where one or both of the boundary spins are not of category $2$ should be treated separately (i.e. in configurations (a), (b), (c), (d) and (e)), because in such configurations the probability distribution of $\alpha$ and $\beta$ are described by $Q^\prime_\alpha$ and $Q^\prime_\beta$ rather than $Q_\alpha$ and $Q_\beta$.
Bearing these two points in mind, 
the probability $P(k,\alpha;l,\beta)$ for each of the configurations presented in \fref{fig:CentralChain} is given by the following expressions, 
\begin{eqnarray}
P(k,\alpha;l,\beta)=
\left\{
\begin{array}{ccc}
Q^\prime_\alpha Q^\prime_\beta&k=l=0&(a)\\
Q^\prime_\alpha Q^\prime_2 Q_2^{l-1}Q_\beta&k=0,~0<l<r&(b)\\
\delta_{\alpha,\beta}Q^\prime_\alpha Q^\prime_2 Q_2^{r-1}&k=0,~l=r&(c)\\
Q^\prime_\beta Q^\prime_2 Q_2^{k-1}Q_\alpha&l=0,~0<k<r&(d)\\
\delta_{\alpha,\beta}Q^\prime_\beta Q^\prime_2 Q_2^{r-1}&l=0,~k=r&(e)\\
{Q^\prime_2}^2 Q_2^{l+k-2}Q_\alpha Q_\beta&k>0,~l>0,~k+l<r&(f)\\
\delta_{\alpha,\beta}{Q^\prime_2}^2 Q_2^{l+k-2} Q_\alpha&k>0,~l>0,~k+l=r&(g)
\end{array}
\right.\label{eq:Pkalb}
\end{eqnarray}
The probability of each configuration is given by a product of probabilities that each relevant spin is of a specific category.
For example, for configuration (g) the quantity ${Q^\prime_2}^2$ refers to the probability of both boundary spins being of category $2$, the power $Q_2^{l+k-2}$ refers to the probability of $l+k-2$ interior spins also being of category $2$ and the term $Q_\alpha$ describes the single interior spin not of category $2$.
The probability of the special case that all spins are of category $2$ is ${Q_2^\prime}^2Q_2^{r-1}$.

The states of spins $0$ and $r$ are determined by $\alpha$ and $\beta$ in \eref{eq:Pkalb}, so that $s_0=S(\alpha)$ and $s_r=S(\beta)$ where,
\begin{eqnarray}
S(\alpha)=\left\{\begin{array}{cc}1,&\alpha=1\\-1,&\alpha = 3\end{array}\right.~.
\end{eqnarray}
Then the expectation value can be calculated by the sum,
\begin{eqnarray}
\left\langle s_0s_r\right\rangle&=&\sum_{\alpha,\beta}\sum_{k=0}^r\sum_{l=0}^{r-k}S(\alpha)S(\beta)P(k,\alpha;l,\beta)+{Q_2^\prime}^2Q_2^{r-1}\label{eq:cfuncSum}
\end{eqnarray}
Substitution of the probabilities given by \eref{eq:Pkalb} into \eref{eq:cfuncSum} gives, 
\begin{eqnarray}
\left\langle s_0s_r\right\rangle&=&(Q_1^\prime-Q_3^\prime)^2+2(Q_1^\prime-Q_3^\prime)Q_2^\prime
\sum_{i=1}^{r-1}Q_2^{i-1}(Q_1-Q_3)\nonumber\\
&+&{Q_2^\prime}^2\sum_{i=2}^{r-1}(i-1)Q_2^{i-2}(Q_1-Q_3)^2
+{Q_2^\prime}^2(r-1)Q_2^{r-2}(1-Q_2)\nonumber\\ 
&+&2Q_2^\prime Q_2^{r-1}(1-Q_2^\prime)+Q_2^\prime Q_2^{r-1}Q_2^\prime~.
\end{eqnarray}

The value of $\langle s_0\rangle=\langle s_r\rangle$ in \eref{eq:CorrFunc} can be found similarly by considering a central chain of infinite length ($r=\infty$) and evaluating the probabilities, $P(k,\alpha)$, that the sites $i$, $0\leq i<k$ ($k \geq 0$) are of category $2$ and that spin $k$ is of category $\alpha$, 
\begin{equation}
P(k,\alpha)=\left\{\begin{array}{cc}Q^\prime_\alpha~,&k=0\\Q^\prime_2Q_2^{k-1}Q_\alpha~,&k>0\end{array}\right.~.
\end{equation}
Therefore the mean magnetisation can be evaluated by a sum,
\begin{eqnarray}
\langle s_0\rangle&=&\sum_\alpha\sum_{k=0}^\infty S(\alpha)P(k,\alpha)\nonumber\\
&=&Q^\prime_1-Q^\prime_3+Q^\prime_2(Q_1-Q_3)\sum_{k=1}^\infty Q_2^{k-1}\nonumber\\
&=&Q^\prime_1-Q^\prime_3+\frac{Q_1-Q_3}{1-Q_2}Q^\prime_2~.
\end{eqnarray}
Using the definition of the correlation function (see \eref{eq:CorrFunc}) it can be found that,
\begin{eqnarray}
C(r)&=&-{{2(Q_1^\prime-Q_3^\prime)(Q_1-Q_3)Q_2^\prime Q_2^{r-1}} \over {1-Q_2}}\nonumber\\
&-&{{{Q_2^\prime}^2(Q_1-Q_3)^2\left[(r-1)(1-Q_2)+Q_2\right]Q_2^{r-2}}\over{(1-Q_2)^2}}\nonumber\\
&+&{Q_2^\prime}^2(r-1)Q_2^{r-2}(1-Q_2)+2Q_2^\prime Q_2^{r-1}(1-Q_2^\prime)+{Q_2^\prime}^2Q_2^{r-1}\nonumber\\
&+&\left[(Q_1^\prime-Q_3^\prime)+{{Q_1-Q_3} \over {1-Q_2}}Q_2^\prime\right]^2-{\langle s_0\rangle}^2~,
\end{eqnarray}
The final two (constant) terms cancel, and the above formula can be rewritten as,
\begin{eqnarray}
C(r)&=&\left\{\frac{4Q_2^\prime Q_1Q_3}{(1-Q_2)^2Q_2}\left[\left(\frac{Q_1^\prime} {Q_1}+\frac{Q_3^\prime}{Q_3}\right)\left(1-Q_2\right)+ \left(2Q_2^\prime-\frac{Q_2^\prime}{Q_2}\right)\right]\right.\nonumber\\
&+&\left.\frac{4Q_1Q_3{Q_2^\prime}^2}{(1-Q_2)Q_2^2}r\right\}\left(Q_2\right)^r~,\label{eq:corrFuncFinalR}
\end{eqnarray}
which completes the derivation of the correlation function given by~\eref{eq:corrFuncFinal}.

\section{Numerical Support}
\label{sec:numerical}

The exact expression~\eref{eq:corrFuncFinal} for correlation function can be supported by means of numerical simulations.
Since
the number of sites in a Cayley tree grows exponentially with the number of generations it contains, a useful numerical description of a system is computationally expensive on this system.
Instead, our numerical calculations have been undertaken on a $q$-regular graph (thin random graph~\cite{Dean2000}), i.e. a graph of
$N$ vertices, in which bonds are placed randomly in such a way that each node has
a coordination number exactly equal to $q$. 
Due to the absence of a boundary, finite size effects are much less pronounced in such a system than in a Cayley tree of the same size.
This graph contains loops, however these loops are small in number, and are irrelevant for large enough systems~\cite{Bollobas2001}. Therefore, a $q$-regular graph gives a good approximation to a Bethe lattice with the same coordination number.
A set of spins is placed on the nodes of the $q$-regular graph. 
These spins interact according to the Hamiltonian
given by~\eref{eq:Hamiltonian} and relax by
Glauber dynamics as described in \sref{sec:Model}.
A similar comparison of numerical data for magnetisation versus external field obtained for a $q$-regular graph with analytical results derived for a Bethe lattice has been performed in~\cite{Dhar1997}.

In our numerical model, each spin in the $q$-regular graph is assigned a random field $h_i$ according to a normal probability distribution,
$h_i\sim {\cal N} (0,\Delta^2)$
with zero mean and variance $\Delta^2$.
All the spins are initally set in the down state, and the system is allowed to evolve according to the dynamics described in \sref{sec:Model} until it is in a stable state.
In order to calculate the correlation function $C(r)$ given by~\eref{eq:CorrFunc}, a mean $\langle s_is_j\rangle$ is calculated for all pairs of spins $i$ and $j$ separated by a chemical distance $r$.
The correlation function is then averaged over several ($10^3$) realisations of disorder.

The $r$-dependence of the correlation function $C(r)$ presented in Fig.~(2) is in good agreement with the exact analytical expression derived in this paper.

\section{Critical Exponents}

\begin{figure}
\begin{center}
\includegraphics[clip=true,width=6cm]{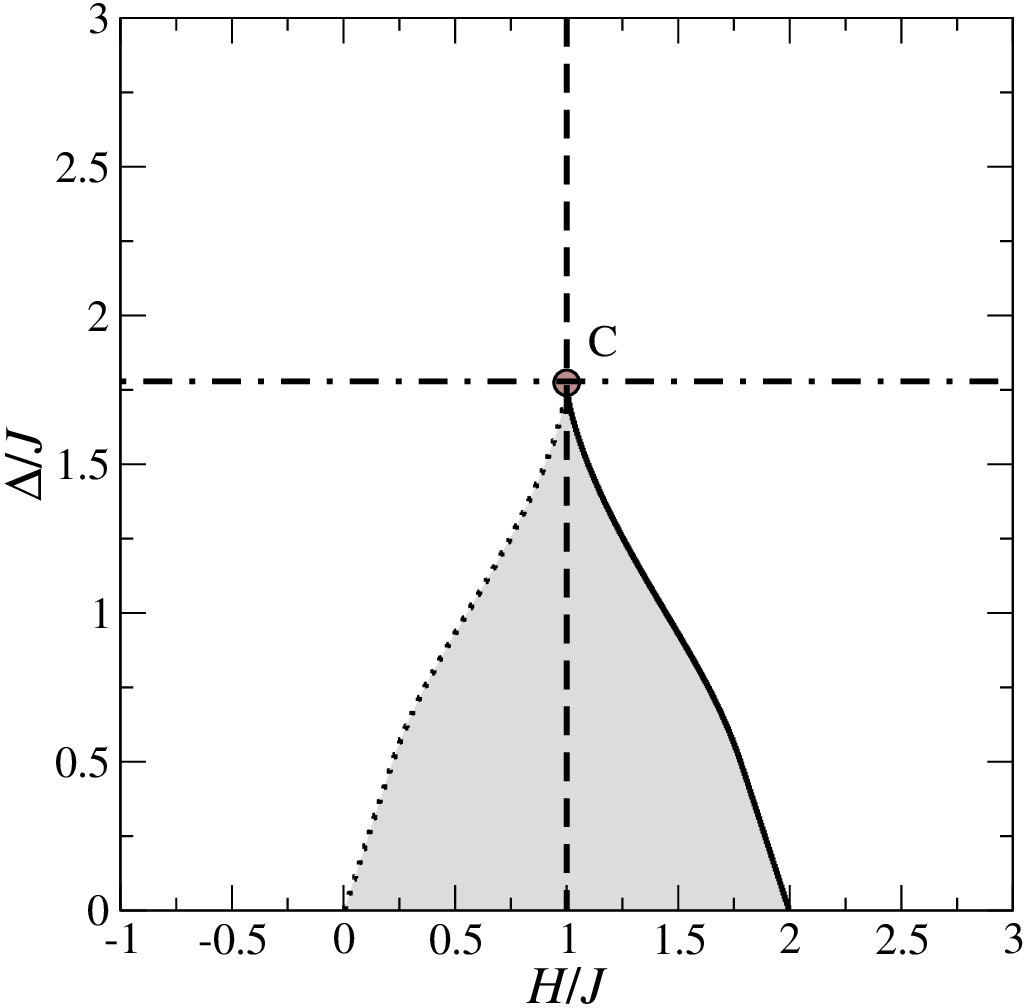}
\end{center}
\caption{\label{fig:PhaseDiagram}
Locus of points in parameter space $(H/J,~\Delta/J)$ for various number of real solutions of the self-consistent equation, $G(P^*)=0$, 
with the shaded and white regions corresponding to three and one solutions, respectively. 
At the boundaries (continuous and dotted lines), the equation $G(P^*)=0$ has two roots, one of which is a multiple root. The values of $(H/J,~\Delta/J)$ marked by the continuous line correspond to the occurrence of the infinite avalanche for increasing external field. 
These loci were obtained for the zt-RFIM defined on a Bethe lattice with $q=4$ with random fields distributed according to a Normal distribution, ${\cal N}(0,\Delta^2)$, in the presence of an external field $H$. 
The critical point at $(H_c/J,~\Delta_c/J)\simeq(1.0,~1.78215895)$ is marked by $C$.
}
\end{figure}

The behaviour of the zt-RFIM can be determined by the number of solutions of the self-consistent equation $G(P^*)=0$ (where $G(P^*)=F(P^*)-P^*$; see~\eref{eq:selfConsistantM}).
In the presence of external field $H$, and with random fields distributed according to a pdf with zero mean and variance $\Delta^2$, the number of solutions of the self-consistent equation depends on the location in parameter space $(H,\Delta)$~\cite{Dhar1997}.
For a Bethe lattice with $q=4$ and a normal distribution of random fields, ${\cal N}(0,\Delta^2)$, the loci of points corresponding to different numbers of real solutions of $G(P^*)=0$ are presented in~\fref{fig:PhaseDiagram}.
The shaded area (excluding the boundaries) corresponds to the range of parameters where the self-consistent equation 
has three real solutions corresponding to different branches of the multivalued function, $\langle m\rangle(H,\Delta)$, for mean magnetisation.
(see the part of the solid curve in~\fref{fig:hysteresis} marked by circles).
The boundaries correspond to the case of two solutions, with the boundary marked by a solid line corresponding to the locus of points where an infinite avalanche occurs as external field increases from $-\infty$ to $\infty$ (see arrow in~\fref{fig:hysteresis}).
Outside the shaded region, the self-consistent equation has a single solution (see dashed line, dot-dashed line and the part of the solid line not marked by circles in~\fref{fig:hysteresis}).
The point at which all three solutions converge is a critical point marked by C in~\fref{fig:PhaseDiagram}, and represents the boundary between low and high disorder, i.e. with and without an infinite avalanche, respectively (see dotted line in \fref{fig:hysteresis}).

In this section, we analyse the behaviour of the correlation length in the zt-RFIM around the critical point $(\Delta,H)=(\Delta_c,H_c)$.
In order to do this, we expand~\eref{eq:xiExpand} in terms of the differences $P^*-P_c$, $H-H_c$ and $\Delta-\Delta_c$ measuring the distance to the critical point, $(P_c,\Delta_c,H_c)$.
The derivative $\partial G/\partial P^*$ in this equation can be expanded in terms of the values $P^*$, $H$ and $\Delta$ near $(P_c,\Delta_c,H_c)$ in the following way,
\begin{eqnarray}
\hskip-70pt
\left.\frac{\partial G}{\partial P^*}\right\vert_{P^*,H,\Delta}&=&\left.\frac{\partial G}{\partial P^*}\right\vert_{P^*_c,H_c,\Delta_c}+
\left.\frac{\partial^2 G}{\partial {P^*}^2}\right\vert_{P^*_c,H_c,\Delta_c}
(P^*-P^*_c)+
\left.\frac{\partial^2 G}{\partial P^*\partial H}\right\vert_{P^*_c,H_c,\Delta_c}
(H-H_c)\nonumber\\
\hskip-70pt
&+&\left.\frac{\partial^2 G}{\partial P^*\partial 
\Delta}\right\vert_{P^*_c,H_c,\Delta_c}
(\Delta-\Delta_c)+
\left.\frac{\partial^3 G}{\partial {P^*}^3}\right\vert_{P^*_c,H_c,\Delta_c}
(P^*-P^*_c)^2~,\label{eq:dGExpand}
\end{eqnarray}
where terms of higher order have been neglected because they do not play an essential role in the vicinity of the critical point. In the above expansion, we have used the fact that the condition for the merging of the three solutions of the self-consistent equation is $\partial G/\partial P^*=\partial^2 G/\partial {P^*}^2=0$.

\begin{figure}
\begin{center}
\includegraphics[clip=true,width=11cm]{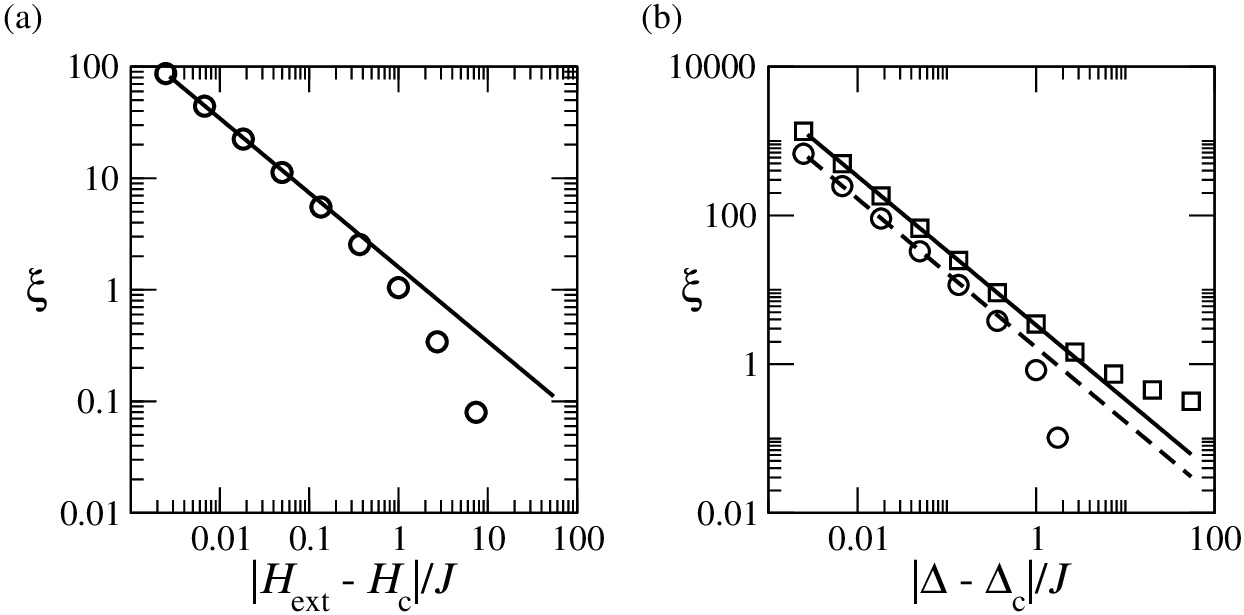}
\end{center}
\caption{\label{fig:CorrLenHDelta} 
Spin-spin correlation length for the zt-RFIM in the case of normally distributed random fields, $h_i\sim {\cal N}(0,\Delta^2)$ {\it vs} (a) external field, $|H-H_c|$, at critical value of disorder ($\Delta=\Delta_c\simeq 1.78215895 J$) and (b) degree of disorder $|\Delta-\Delta_c|$ at critical value of external field ($H=H_c=J$).
In panel (a), the behaviour on both sides of criticality is identical (it depends only on $|H-H_c|$), and is shown by the circles. 
In panel (b), the behaviour is different above and below criticality, and both behaviours are shown (squares for $\Delta>\Delta_c$ and circles for $\Delta<\Delta_c$).
The solid line in panel (a) and the solid and dashed lines in panel (b) are plots of~\eref{eq:dGH} and~\eref{eq:dGdelta}, showing convergence of the exact correlation length to these equations near criticality.
}
\end{figure}

The observed critical point is known to be a saddle-node transition~\cite{Dhar1997}, meaning that the solution, $P^*$, 
of~\eref{eq:selfConsistantM} exhibits 
the known mean-field exponents 
around this point.
Explicitly, the behaviour of $P^*-P_c$ around criticality with fixed $\Delta=\Delta_c$ can be obtained by following the standard method for saddle-point transitions (see e.g.~\cite{Chaikin_Lubensky_2000}), i.e. expanding $G(P^*)=0$ along the dot-dashed line in \fref{fig:PhaseDiagram}  in terms of $P^*-P_c$ and $H-H_c$. 
Keeping terms only to lowest order in $H-H_c$ and $P^*-P_c$, the equation reduces to the following expression,
\begin{equation}
P^*-P_c=\pm\left(6{\left.\frac{\partial G}{\partial H}\right|_{P_c,H_c,\Delta_c}}\right)^{1/3}
          \left(-\left.{\frac{\partial^3 G}{\partial {P^*}^3}}\right|_{P_c,H_c,\Delta_c}\right)^{-1/3}\left|H-H_c\right|^{1/\delta}\label{eq:expdelta}~,
\end{equation}
where $\pm$ refers to the sign of $H-H_c$ and the exponent takes its mean-field value, $\delta=3$.
Similarly, expanding $G(P^*)=0$ along the dashed line in \fref{fig:PhaseDiagram} 
and keeping lowest order terms in $\Delta-\Delta_c$ and $P^*-P_c$, the behaviour is found to be,
\begin{eqnarray}
\hskip-70pt
P^*-P_c=\left\{
\begin{array}{cc}
-\left(6{\left.{\frac{\partial^2 G}{\partial\Delta\partial P^*}}\right|_{P_c,H_c,\Delta_c}}\right)^{1/2}
\left(-\left.{\frac{\partial^3 G}{\partial {P^*}^3}}\right|_{P_c,H_c,\Delta_c}\right)^{-1/2}
\left|\Delta-\Delta_c\right|^{\beta},
& \Delta<\Delta_c\\
0,
& \Delta\geq\Delta_c 
\end{array}\right.
\label{eq:expbeta}
\end{eqnarray}
where the exponent also takes its mean-field value, $\beta=1/2$.
The substitution of~\eref{eq:expdelta} and~\eref{eq:expbeta} into~\eref{eq:dGExpand} results in,
\begin{eqnarray}
\xi=-\left(\left.\frac{\partial G}{\partial P^*}\right\vert_{P^*,H,\Delta_c}\right)^{-1}&=&
\frac{\sqrt[3]{6}}{3}\left(-\frac{\partial^3G}{\partial {P^*}^3}\right)^{-\frac{1}{3}}\left({\frac{\partial G}{\partial H}}\right)^{-\frac{2}{3}}\left|H-H_c\right|^{-\mu}~,\label{eq:dGH}
\\
\xi=-\left(\left.\frac{\partial G}{\partial P^*}\right\vert_{P^*,H_c,\Delta}\right)^{-1}&=&
\cases{
\left|{\left.{\frac{\partial^2G}{\partial\Delta\partial P^*}}\right.}\right|^{-1}\left|\Delta-\Delta_c\right|^{-\nu},\Delta>\Delta_c\\
\frac{1}{2}\left|{\left.{\frac{\partial^2G}{\partial\Delta\partial P^*}}\right.}\right|^{-1}\left|\Delta-\Delta_c\right|^{-\nu},\Delta<\Delta_c~,
}\label{eq:dGdelta}
\end{eqnarray}
where the critical exponents for the correlation length take the values, $\nu=1$ and $\mu=2/3$. As remarked in \sref{sec:expand}, these values are different from the mean-field exponents, $\nu_{\rm{MF}}=1/2$ and $\mu_{\rm{MF}}=1/3$.
In order to confirm the values of $\nu$ and $\mu$, we have calculated the 
spin-spin correlation length for zt-RFIM in the case of normally distributed random fields, $h_i\sim {\cal N}(0,\Delta^2)$ by solving~\eref{eq:Fm} and~\eref{eq:selfConsistantM} numerically and using~\eref{eq:Q}~and~\eref{eq:xi}.
The results for the correlation length {\it vs} $H-H_c$ (for fixed $\Delta=\Delta_c\simeq 1.78215895 J$) and $\Delta-\Delta_c$ (for fixed $H=H_c=J$) are compared with~\eref{eq:dGH}~and~\eref{eq:dGdelta} in \fref{fig:CorrLenHDelta}(a) and \fref{fig:CorrLenHDelta}(b), respectively.
A similar check has been performed up to $q=10$, finding no deviation of the exponents from that for $q=4$.

\end{document}